\documentclass[prb,aps,showpacs]{revtex4}
\usepackage{graphicx,psfig}

\usepackage{color}

\newcommand{\beq}{\begin{equation}}
\newcommand{\beqa}{\begin{eqnarray}}
\newcommand{\eeq}{\end{equation}}
\newcommand{\eeqa}{\end{eqnarray}}

\begin{document}

\title{Surface states, Friedel oscillations, and spin accumulation in $p$-doped semiconductors}

\author{Tudor D. Stanescu}
\affiliation{Department of Physics, University of Virginia,
Charlottesville, VA 22904-4714}
\author{Victor Galitski}
\affiliation{Department of Physics, University of Virginia,
Charlottesville, VA 22904-4714}
\date{\today}

\begin{abstract}
We consider a hole-doped semiconductor with a sharp boundary
 and study the boundary spin accumulation in response to a charge current.
 First, we solve exactly a single-particle quantum mechanics problem described
by the isotropic  Luttinger model in half-space and construct an
orthonormal basis for the corresponding Hamiltonian. It is shown
that the complete basis includes two types of eigenstates. The first
class of states contains conventional incident and reflected waves,
while the other class includes localized surface states.  Second, we
consider a many-body system in the presence of a charge current
flowing parallel to the boundary. It is shown that the localized
states contribute to spin accumulation near the surface. We also
show that the spin density exhibits current-induced Friedel
oscillations with three different periods determined by the Fermi
momenta of the light and heavy holes. We find an exact asymptotic
expression for the Friedel oscillations far from the boundary. We
also calculate numerically the spin density profile and compute the
total spin accumulation, which is defined as the integral of the
spin density in the direction perpendicular to the boundary. The
total spin accumulation is shown to fit very well the simple formula
$S_{\rm tot} \propto \left( 1 - m_L/m_H \right)^2$, where $m_L$ and
$m_H$ are the light- and heavy-hole masses. The effects of disorder
are discussed. We estimate the spin relaxation time in the Luttinger
model and argue that spin physics cannot be described within the
diffusion approximation.
\end{abstract}

\maketitle

\section{Introduction}

Hole-doped semiconductors are a very well studied and industry
developed class of materials. The fundamental description of these
materials is usually based on effective models such as the Kane
model or Luttinger model,~\cite{Lutt} which capture most of the
properties of a semiconductor. A key ingredient of these models is
spin-orbit interaction, which couples  the momentum with the orbital
and spin degrees of freedom. It should be noted that the latter
degrees of freedom in semiconductor systems attracted attention only
very recently, when it was recognized that the spin-orbit coupling
may lead to the possibility of spin control by electric
means.~\cite{Zhang_Science,Sinova_etal} On one hand, the predicted
spin-charge coupling opens a possibility for new useful spintronics
applications.~\cite{Awschalom,ZFDS} On the other hand, it leads to a
variety of new theoretical problems, which need to be clarified in
order to understand the relevant
experiments\cite{Kato,Awschalom2,Wunderlich}. The theoretical
description  of the intrinsic spin-Hall
effect,~\cite{BS,SZhang,Sinova_etal} which is
one of the spin-charge coupling phenomena, relies on an elegant
mathematical structure known as the Fermi surface Berry's
phase.~\cite{Sundaram_Niu,zhang,Haldane_AHE} This structure
originates from the spin-orbit splitting of the bands. Band
crossings become sources of a fictitious magnetic field in momentum
space, which leads to a non-trivial Berry's phase and may affect
certain observables.  In particular it leads to an anomalous
contribution to the velocity operator. Consequently, any
 quantum mechanical operator, which involves the velocity
acquires an anomalous contribution. An important example is the spin
current operator (usually defined as a symmetrized product of the
spin density and the velocity), which may acquire an
anomalous component as well, if an electric current is
present.~\cite{Sinova_etal,EHR,TseDS,Orbitronics,BZ_holegas,Schl_Loss,Bernevig_Zhang}
The existence of the anomalous contribution to the spin current
perpendicular to the electric current is an important prediction of
the spin Hall effect theory. However, a direct experimental check of
this prediction, while possible in principle, is not straightforward
because the spin is not conserved, and as such, the spin current has
no obvious physical meaning.~\cite{Niu,Rash1} Alternatively one can
experimentally probe current-induced equilibrium spin density and
that is what usually is measured in experiment. It is therefore
desirable to develop a theory, which would allow one to calculate
observables such as boundary spin accumulation directly
 and provide a clear understanding of the physical processes behind this effect. It is also desirable to search for other
possible manifestations of the Fermi surface Berry's phase apart from the anomalous velocity.

In this paper we consider a three-dimensional hole-doped
semiconductor described by the isotropic Luttinger model and in the
presence of a boundary. We mostly discuss the clean limit when no
impurities are present or, alternatively,  a disordered system but
only at small (ballistic) length-scales. We note here that
 the application of an electric field to a perfectly clean system
would result in a non-equilibrium state and time-dependent spin
density. To access equilibrium spin-Hall physics we assume that
either the voltage drop occurs only in the contacts,  or that
there is a
relaxation provided by impurities. In both cases there is an
equilibrium charge current, which determines spin accumulation near
the boundary.

Evaluation of the current-induced spin density involves solving the
single-particle Schr{\"o}dinger equation for the Luttinger
Hamiltonian in  a half-space. The corresponding solution  contains a
few important features, which are quite different from the usual
single-band quantum-mechanical problem. While the states of the bulk
Luttinger model can be classified by quantum numbers corresponding
to the double-degenerate heavy and light-hole bands, the boundary
does not conserve these quantum numbers and mixes up different
bands. For example, if a heavy-hole with positive chirality
propagates towards the boundary in the direction close to normal, it
gets reflected in all bands, which are heavy- and light-holes with
positive and negative chiralities, and the reflected light holes
have the angle of reflection (measured from the normal to the
surface) larger  than the angle of incidence. An
important property of the solution is that for large enough angles
of incidence, light holes are not reflected from the boundary at all,
but instead get localized near the surface. These states are similar
to the Tamm states,~\cite{Tamm} which appear in crystals due to an
abrupt change of the electronic band structure at the boundary. We
also point out that one can draw an intuitive analogy with
 optics by imagining that the heavy holes
occupy a half-space  with  high index of refraction,
while the light holes occupy the other half with a lower index of
refraction. A wave propagating say from the medium with the high
index of refraction (heavy hole) may get reflected from the
interface (remains a heavy hole) or may get refracted and propagate
in the other medium (becomes a light hole). For large enough angles
of incidence, one expects total internal reflection, which is
somewhat similar to the appearance of localized light-hole states in
our language.
It can be shown  that these localized light-hole states
contribute to spin accumulation if a current is present
(below we study accumulation of the total orbital momentum,
but occasionally call it ``spin accumulation'' for the sake of
brevity).

To qualitatively understand the physics of the boundary spin
accumulation in a many-particle system it is useful to recall the
well-known problem of a free Fermi gas in the presence of a
boundary. The boundary (or any other perturbation for this matter)
leads to the Friedel oscillations in the particle density with the
period of two Fermi momenta. We note that the integral of
the density over distance reduces to the bulk density, {\em i.e.},
there is no boundary charge accumulation since the latter is
conserved. A similar oscillatory behavior  of the spin density may
occur in spin-orbit coupled systems ({\em e.g.}, described by the
Luttinger model) if a current is present. There are, however, two
important differences: (i)~The existence of multiple periods of
Friedel oscillations (or
beatings) due to two distinct Fermi momenta corresponding to the
light and
heavy holes; (ii)~The non-zero integral of the spin density due to
the non-conservation of spin. Summarizing the above qualitative
discussion, we argue that non-zero spin density, which
 appears near the boundary in response to an applied current can be viewed
 as current-induced Friedel oscillations with non-zero spin
 accumulation originating from the localized (Tamm-like) surface
 states.

Our paper is structured as follows: In Sec.~\ref{Sec:Eignestates} we
solve the quantum-mechanical problem of a particle described by the
Luttinger Hamiltonian in a half-space. We show that for the angles
of incidence greater than a critical angle $\theta_c$, the
eigenstates  contain modes  that are localized in the direction
normal to the boundary.
We construct a full orthonormal basis for the problem.
Even though the formulation of the problem is very straightforward,
its solution is technically challenging due to a cumbersome matrix
structure of the Hamiltonian. Subsections \ref{Sec:Orthogonality}
and \ref{Sec:Counting} are devoted to a formal proof that certain
symmetrized combinations of incident and reflected waves constitute
a full orthonormal basis for the Hamiltonian. A reader not
interested in the formal proof should skip to
Sec.~\ref{Sec:Many-body}.

In Sec.~\ref{Sec:Many-body} we address the question of boundary spin
accumulation in response to an external current. Using the
orthonormal basis constructed in Sec.~\ref{Sec:Eignestates}, we
obtain numerically the spin density profile near the boundary for
various values of the ratio of the light and heavy hole masses, $\xi
= m_L/m_H$. We also extract analytically the large-distance
asymptotic behavior of the spin density, which is shown to oscillate
with three distinct periods
$2 k_{\rm F}^{(H)}$, $2 k_{\rm F}^{(L)}$, and
$k_{\rm F}^{(H)} +  k_{\rm F}^{(L)}$
(where $k_{\rm F}^{(H)}$ and $k_{\rm F}^{(L)}$
are Fermi momenta of the heavy and light holes correspondingly) and
decay away from the boundary as a power law, $\propto 1/r^2$. In
subsection \ref{Sec:Accumulation}, we define a quantity, which we
call spin accumulation, by integrating the spin density over
distance in the direction perpendicular to the boundary. We obtain the
spin accumulation numerically and show that the total spin
accumulation fits  the simple formula $S_{\rm tot} \propto \left( 1
- \xi \right)^2$.

In Sec.~\ref{Sec:Disorder}, we discuss qualitatively the effects of
disorder on the current-induced Friedel oscillations. Using the
analogy between the spin accumulation effect and the usual Friedel
(or RKKY)
oscillations in disordered metals, we argue that one should expect
an interesting behavior of the current-induced Friedel oscillations
in the spin density in a disordered system. We  argue that
while the system-wide average value of the boundary spin density
decays exponentially away from the boundary as $e^{-r/l}$ (where $l$
is the mean free path), the higher moments and the {\em typical}
spin density still decays as a power law $r^{-2}$. However, the
latter has a random sign and the spin accumulation ({\em i.e.}, spin
density averaged over large enough distances) decays exponentially
as $e^{-r/L_s}$ (where $L_s$ is a spin relaxation length). We
calculate the spin relaxation time and show that for all reasonable
values of the spin-orbit coupling, the spin relaxation time is very
short and  has a universal value $\tau_s = 3 \tau/2$. This  short
spin relaxation time implies that the spin relaxation length is
 of the order of the mean free path and, therefore, the
hydrodynamic diffusion approximation does not apply for the
Luttinger model.

\section{Eigenstates for the Luttinger Hamiltonian in half-space}
\label{Sec:Eignestates}

The physics of a hole-doped semiconductor with diamond or
zinc-blende structure is often described by  the effective Luttinger
Hamiltonian\cite{Lutt}: \beq \hat{\cal H}_{\rm Lut} =
\frac{1}{2m}\left[\left(1 + \frac{5}{2}\gamma \right){\bf k}^2
-2\gamma ({\mathbf k}\cdot \hat{\mathbf S})^2\right],
\label{luttham} \eeq where m is the effective mass, ${\bf k} =
(k_x,k_y,k_z)$ is the momentum and $\hat{\bf S} = (\hat S_x,\hat
S_y,\hat S_z)$ represents the total angular momentum $3/2$ of the
atomic orbital, i.e. the sum of the orbital angular momentum and the
spin. The total angular momentum can be represented by three
4$\times$4 matrices with explicit expressions given in Appendix
\ref{AppA}. For simplicity, we consider the spherically symmetric
model described by one Luttinger parameter\cite{Lutt}, $\gamma$, and
we choose the units so that $\hbar=1$.

For a system with translational symmetry, the Hamiltonian
(\ref{luttham}) can be diagonalized in a basis in which the helicity
operator $\lambda = {\bf k}\cdot\hat {\bf S}/k$ is also
diagonal. For a given wave vector ${\bf k}$, $\hat{\cal H}_{\rm
Lut}$ has two double degenerate eigenvalues
\beqa
\epsilon_H({\bf k}) &=& \frac{1-2\gamma}{2m}k^2 \equiv \frac{k^2}{2m_H} ~~~~~~~~~~~~~~~~\mbox{for} ~~~~\lambda= \pm \frac32, \nonumber \\
\epsilon_L({\bf k}) &=& \frac{1+2\gamma}{2m}k^2 \equiv
\frac{k^2}{2m_L} ~~~~~~~~~~~~~~~~\mbox{for} ~~~~\lambda= \pm
\frac12.
\eeqa
These two bands are referred to as  heavy-holes and
light-holes, respectively. The corresponding eigenfunctions can be
expressed in terms of a four-component spinor,
\beq \Phi_{{\bf
k}\lambda}({\bf r}) = \frac{1}{\sqrt{\Omega}}e^{i{\bf k}{\bf r}}
~U_{\lambda}({\bf n_k}),   \label{spinor}
\eeq
where $\Omega$ is the
volume of the system, the label $\lambda$ is $(\pm 3/2) \equiv
(H\pm)$ for the heavy-holes and $(\pm 1/2) \equiv (L\pm)$ for the
light-holes, and the spinors $U_{\lambda}({\bf n_k})$ depend on the
orientation of the wave vector, ${\bf n_k} = {\bf k}/ |{\bf k}|$.
Their explicit forms are given in Appendix \ref{AppA}.

Next we consider a similar problem for a system defined in
 half-space, i.e. in the presence of a sharp boundary in the z
direction defined by the potential
\beq
V({\mathbf r}) = \left\{
\begin{array}{ll}0 ~~~~ \mbox{if $z > 0$} \\ \infty ~~\mbox{if $z < 0$}\end{array}\right.
\eeq
\begin{figure}
\begin{center}
\includegraphics[width=0.5\textwidth]{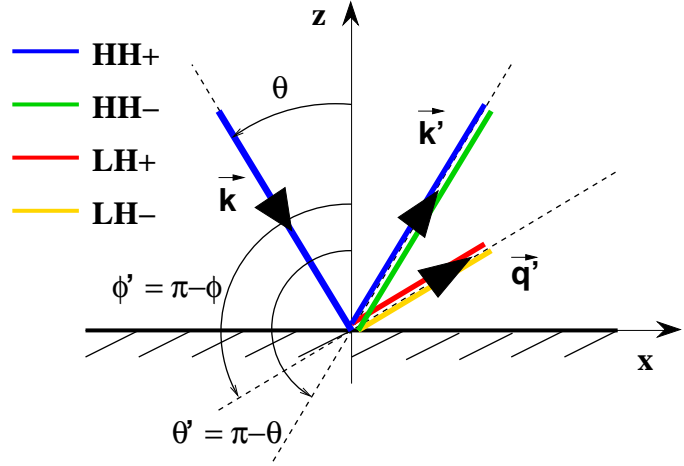}
\caption{(Color online) Scattering of an incident heavy-hole plane
wave. By convention, the coordinate system is chosen so that $k_y=0$
and  $k_x \ge 0$. The angles $\theta$, $\theta^{\prime}$ and
$\varphi$ that describe the direction of
propagation for a given wave are defined as the angles between the
z-axis and $-{\bf p}$, where ${\bf p} \in \{{\bf k, ~k^{\prime},
~q^{\prime}}\}$, and belong to the interval $[0, \pi]$. The regime
represented here corresponds to $\theta < \theta_c$ (see text) and
the incident and reflected waves are described by the wave-functions
defined by Eq.~(\ref{spinor}).} \label{FIG1}
\end{center}
\end{figure}
Solving the quantum problem described by the Luttinger
Hamiltonian~(\ref{luttham}) in the presence of the boundary is still
a straightforward exercise. However, finding an orthonormal basis
for this system, represents a technical  challenge because of two
main reasons: 1) The new eigenstates are, in general, linear
combinations of bulk eigenfunctions $\Phi_{{\bf k}\lambda}({\bf r})$
and are not automatically orthogonal on each other, and 2) The
proper counting of modes is difficult, as the standard technique of
imposing periodic boundary conditions and discretizing the momentum
is not immediately applicable for eigenstates involving combinations
of heavy and light holes. In this Section, we will direct our effort
toward solving these difficulties with the goal of constructing an
orthonormal basis for the Luttinger Hamiltonian in a half space. Let
us notice that while $k_x$ and $k_y$ are still good quantum numbers
for the full Hamiltonian $\hat{\cal H} =\hat{\cal H}_{\rm Lut} +
V({\bf r})$, $k_z$ and the helicity $\lambda$ are not. The wave
functions have to vanish at the boundary, i.e. for $z=0$, while for
positive values of z they are superpositions of incident and
reflected waves. Let us consider the case of a $\lambda=+3/2$
incident heavy-hole (see Fig.~\ref{FIG1}). We can always choose the
coordinate system so that $k_y=0$ and $k_x \ge 0$ by properly
rotating the axes.
The full wave function corresponding to the reflection process
represented schematically in Fig.~\ref{FIG1} has the following form
 for positive values of z,
\beqa \psi_{\bf k}^{(H+)}({\bf
r}) = \frac{e^{ik_x x}}{\sqrt{C}} \left\{U_{H+}(\theta)e^{ik_z
z}\right. &+& A_1~ U_{H+}(\pi-\theta)e^{-ik_z z} + A_2~
U_{H-}(\pi-\theta)e^{-ik_z z}
 \nonumber \\
&+& \left. B_1~ U_{L+}(\pi-\phi)e^{i q_z z} + B_2~
U_{L-}(\pi-\phi)e^{i q_z z}\right\},            \label{psihp} \eeqa
where C is a normalization factor. All the reflection coefficients
$A_i(\theta)$ and $B_i(\theta)$ are in general non-zero and  are
determined by the boundary condition $\psi_{\bf k}(z=0) = 0$. The
parameters that describe the reflected waves can be  determined from
the parameters of the incident wave using the momentum and energy
conservation laws. Explicitly we have for the heavy-holes \beq
\left\{
\begin{array}{lll}
k_x^{\prime} = ~~k_x;\\
k_y^{\prime} = ~~k_y \equiv 0; \\
k_z^{\prime} = -k_z;
\end{array}
\right.  ~~~~~ \rightarrow ~~~~~ \theta^{\prime} = \pi-\theta,  \label{ttheta}
\eeq
where  $\theta = \arccos(-k_z/k) \in [0,\pi/2]$ is the angle of the
incident heavy-hole. Similarly we obtain for the light-holes
\beq
\left\{
\begin{array}{lll}
q_x^{\prime} = k_x;\\
q_y^{\prime} = k_y \equiv 0; \\
q_z^{\prime} = k[\xi - \sin^2(\theta)]^{1/2};
\end{array}
\right.  ~~~~~ \rightarrow ~~~~~ \phi^{\prime} = \pi-\phi = \pi -
\arccos\left(\sqrt{1-\frac{\sin^2(\theta)}{\xi}}\right),
\label{pphi} \eeq where $\xi = m_L/m_H \equiv
(1-2\gamma)/(1+2\gamma)$ is the ratio between the light-hole and
heavy-hole masses. Consequently, the wave function (\ref{psihp}) is
an eigenfunction of the full Hamiltonian with an eigenvalue
$\epsilon_{\bf k} = k^2/2m_H = q^2/2m_L$. This solution exists as
long as $q_z$ is a real number, i.e. for incident angles smaller
that the critical angle \beq \theta_c = \arcsin(\sqrt{\xi}). \eeq
For $\theta > \theta_c$ there are no light-holes  propagating in the
z-direction but, instead, the scattering problem has solutions that
are localized at the boundary. This situation is illustrated
schematically in Fig.~\ref{FIG1a}.
\begin{figure}
\begin{center}
\includegraphics[width=0.5\textwidth]{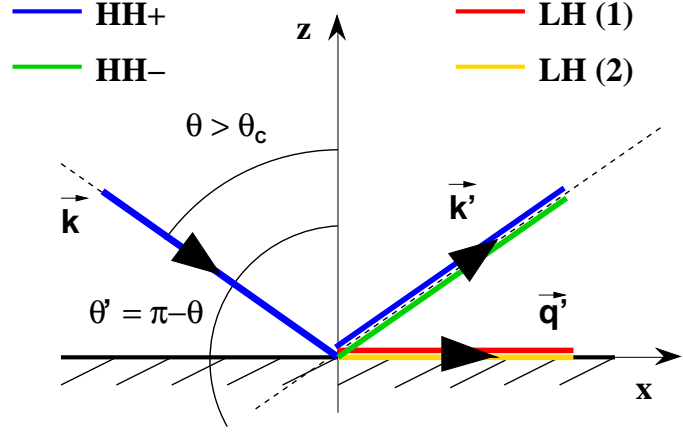}
\caption{(Color online) Scattering of an incident heavy-hole plane
wave with the incident angle $\theta > \theta_c$. The reflected
light-holes propagate parallel to the boundary and are localized in
the z-direction. The wave function describing the reflection process
is given by Eq.~(\ref{psihp2}). Notice that the localized modes are
described by two independent spinors $V_{1}(\chi)$ and $V_{2}(\chi)$
(see Appendix~\ref{AppA}).} \label{FIG1a}
\end{center}
\end{figure}
The localized
modes are characterized by an imaginary wave vector $q_z = iQ$
and the corresponding wave function becomes
\beqa
\psi_{\bf k}^{(H+)}({\bf r}) = \frac{e^{ik_x x}}{\sqrt{C}} \left\{U_{H+}(\theta)~e^{ik_z z}\right. &+& A_1~ U_{H+}(\pi-\theta)~e^{-ik_z z} + A_2~ U_{H-}(\pi-\theta)~e^{-ik_z z}
 \nonumber \\
&+& \left. B_1~ V_{1}(\chi)~e^{-Q z} + B_2~ V_{2}(\chi)~e^{-Q z}\right\},            \label{psihp2}
\eeqa
where $V_{i}(\chi)$ are spinors describing the evanescent modes
and the reflection  coefficients are determined, as before, by the
boundary
condition. Explicit expressions for the $V_{i}(\chi)$ spinors are
given in Appendix \ref{AppA}.
The angle $\chi$ and the wave vector that characterize  the
localized states are determined again using the conservations laws
and we have
\beq
\left\{
\begin{array}{lll}
q_x^{\prime} = k_x;\\
q_y^{\prime} = k_y \equiv 0; \\
q_z^{\prime} \equiv iQ =  ik[\sin^2(\theta) - \xi]^{1/2};
\end{array}
\right.  ~~~~~ \rightarrow ~~~~~ \chi =
\arccos\left(\frac{\sqrt{\xi}}{\sin(\theta)}\right),   \label{pphi2}
\eeq Notice that all the modes contained in the superpositions that
define the wave-function in Eq.~(\ref{psihp}) and Eq.~(\ref{psihp2})
are eigenfunctions of the Luttinger Hamiltonian corresponding to the
same eigenvalue, \beqa \hat{\cal H}_{\rm Lut}
\left[U_{H\pm}(\theta)~e^{i{\bf kr}}\right] = \epsilon
_k\left[U_{H\pm}(\theta)~e^{i{\bf kr}}\right], ~~~~&~&~~~~\hat{\cal
H}_{\rm Lut} \left[U_{L\pm}(\phi)~e^{i(k_x x + q_z z)}\right] =
\epsilon _k\left[U_{L\pm}(\phi)~e^{i(k_x x + q_z z)}\right], \nonumber \\
\hat{\cal H}_{\rm Lut} \left[V_{i}(\phi)~e^{ik_x x - Q z}\right] &=&
\epsilon _k\left[V_{i}(\phi)~e^{ik_x x - Q z)}\right], \eeqa where
\beq \epsilon_k = \frac{k_x^2+k_z^2}{2m_H} ~=~
\left\{\begin{array}{lll} \frac{k_x^2+q_z^2}{2m_L},
~~~~~~~~~~~~~\mbox{if}~~\theta < \theta_c,
\\
~\\
\frac{k_x^2-Q^2}{2m_L}, ~~~~~~~~~~~~\mbox{if}~~\theta > \theta_c.
\end{array}\right.
\eeq Also notice that all the scattering angles and the wave vectors
can be expressed uniquely in terms  of $\theta$ and
$k=(k_x^2+k_y^2+k_z^2)^{1/2}$. In addition, the scattering
coefficients $A_i$ and $B_i$ in Eq.~(\ref{psihp}) and
Eq.~(\ref{psihp2}) are functions of
the angle  $\theta$ of the incident heavy-hole. Their explicit form
is given in Appendix~\ref{AppB}.

\begin{figure}
\begin{center}
\includegraphics[width=0.6\textwidth]{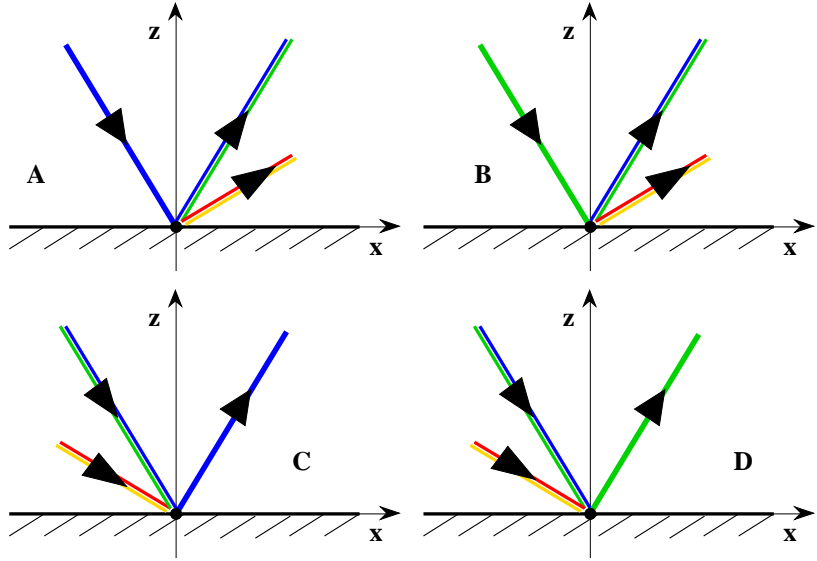}
\caption{(Color online) Family of reflection processes characterized
by a given pair of reflection  angles ($\theta$, $\phi(\theta)$). The
color code is the same as in Fig.~\ref{FIG1}. Panel A represents
schematically the same event as Fig.~\ref{FIG1}, the scattering of a
heavy-hole with helicity $\lambda = +3/2$, while panel B represents
the scattering of an incident heavy-hole with opposite helicity,
$\lambda = -3/2$. Panels C and D illustrate ``re-combination''
processes where superpositions of heavy-holes and light-holes are
reflected into a single heavy-hole mode with helicity $+3/2$ and
$-3/2$, respectively. For incident heavy-hole angles
$\theta>\theta_c$, the light-holes propagate parallel to the
boundary ($\phi = \pi/2$), while they are  localized in the
z-direction, and are described by the spinors $V_i(\chi)$ given by
Eq.~(\ref{Vqchi}).} \label{FIG2}
\end{center}
\end{figure}
\begin{figure}
\begin{center}
\includegraphics[width=0.6\textwidth]{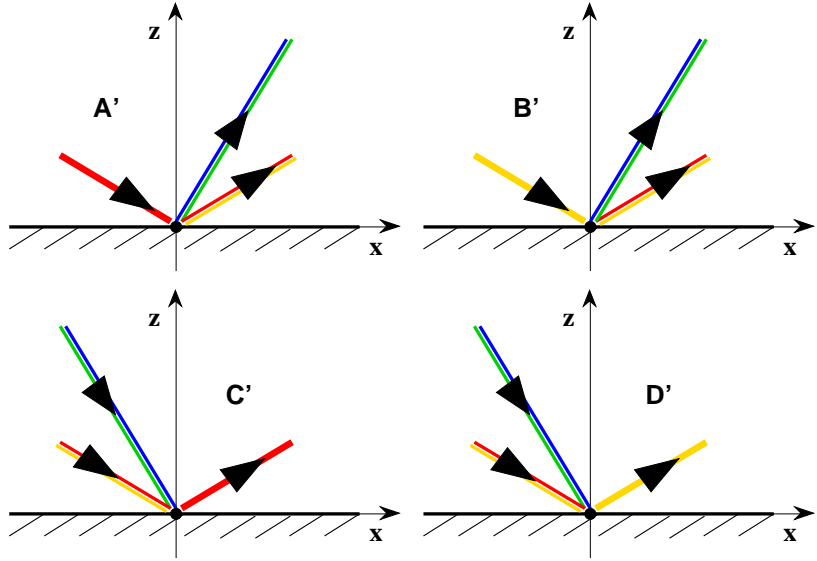}
\caption{(Color online) Light-hole reflections that belong to the
same family as the processes  represented schematically in
Fig.~\ref{FIG2}. These reflections involve a single incident (panels
A' and B') or reflected (panels C' and D') light-hole mode. The
corresponding wave-functions can be written as linear combinations
of wave-functions describing the heavy-hole scattering processes
represented in Fig.~\ref{FIG2}. Same color code as in Fig.
\ref{FIG1}: blue - (HH+), green - (HH-), red - (LH+), yellow -
(LH-). } \label{FIG3}
\end{center}
\end{figure}

The wave function (\ref{psihp}) is not the only eigenstate of the
Hamiltonian $\hat{\cal H} =\hat{\cal H}_{\rm Lut}+V({\bf r})$ with
the eigenvalue $\epsilon_k = k^2/2m_H$ that can be expressed as a
superposition of propagating modes characterized by the scattering
angles ($\theta$, $\phi(\theta)$). For example, a similar state can
be obtained by starting  with an incident heavy-hole with helicity
$\lambda=-3/2$, (HH-). Moreover, we can imagine ``re-combination''
events, such as those represented schematically in Fig.~\ref{FIG2}
(panels C and D), or the corresponding scattering processes of
light-holes shown in Fig.~\ref{FIG3}. From this family of states one
can extract sets of four linearly independent eigenfunctions, while
all the others states can be expressed as linear combinations of
these vectors. Our task is to identify such a complete set of
orthogonal wave functions that would allow us to construct an
orthonormal basis for the full quantum problem. The basic idea is to
construct this set by taking certain symmetric linear combinations
of the eigenfunctions represented in Fig.~\ref{FIG2}. The general
expression for such a symmetrized eigenstate is
\beqa
\Psi_{\bf k}^{\alpha}({\bf r}) &=& \frac{1}{\sqrt{\Omega}}e^{ik_x x}\left\{\left[c_1^{\alpha}~U_{H+}(\theta)+ c_2^{\alpha}~U_{H-}(\theta)\right]e^{ik_z z}\right. + \left[c_3^{\alpha}~U_{L+}(\phi)+ c_4^{\alpha}~U_{L-}(\phi)\right]e^{iq_z z}\nonumber \\
&+& \left.\left[c_5^{\alpha}~U_{H+}(\pi-\theta) +
c_6^{\alpha}~U_{H-}(\pi-\theta)\right]e^{-ik_z z} +
\left[c_7^{\alpha}~U_{L+}(\pi-\phi) +
c_8^{\alpha}~U_{L-}(\pi-\phi)\right]e^{-iq_z z}\right\},
\label{psisymm}
\eeqa
where $|c_i^{\alpha}|^2 =
|c_{i+4}^{\alpha}|^2$, $i\in\{1, 2, 3, 4\}$, i.e. the weight of each
mode is the same for both the incident and the reflected waves. In
constructing the symmetrized eigenfunctions, we use the fact that
the reflection coefficients for the family of states described in
Fig.~\ref{FIG2} can be expressed in terms of the reflection
coefficients of the state $\psi^{(H+)}\equiv\psi^A$ given by
Eq.~(\ref{coeff}). We write  the  corresponding wave-functions in the
most general form given by Eq. (\ref{psisymm}). For example, the
wave-function $\psi_{\bf k}^A$, corresponding to the process
represented in Fig. \ref{FIG2}A, will have the coefficients
$c_1^A = 1$, $c_2^A = c_3^A = c_4^A = 0$, $c_5^A = A_1$, etc., up to
an overall normalization factor that we omit.
In Table \ref{TBL1} we summarize the relations between the
coefficients of the wave-functions corresponding to the reflection
processes represented in Fig. \ref{FIG2}, omitting an overall
normalization factor.
\begin{table}[h!b!p!]
\begin{tabular}{|c||c|c|c|c||c|c|c|c|}
\hline
~ & $c_1^{\alpha}$ & $c_2^{\alpha}$ & $c_3^{\alpha}$ & $c_4^{\alpha}$ & $c_5^{\alpha}$ & $c_6^{\alpha}$ & $c_7^{\alpha}$ & $c_8^{\alpha}$ \\
\hline \hline
$\psi^A$ & 1 & 0 & 0 &  0 & $A_1$ & $A_2$ & $B_1$ & $B_2$ \\
\hline
$\psi^B$ & 0 & 1 & 0 &  0 & $-A_2$ & $A_1$ & $B_2$ & $-B_1$ \\
\hline
$\psi^C$ & $A_1$ & $-A_2$ & $-B_1$ & $B_2$ & 1 & 0 & 0 &  0 \\
\hline
$\psi^D$ & $A_2$ & $A_1$ & $B_2$ & $B_1$ & 0 & 1 & 0 &  0 \\
\hline
\end{tabular}
 \caption{Coefficients for the  family of wave-functions  represented in Fig.~\ref{FIG2}. The wave functions are expressed in the general form given by Eq.~(\ref{psisymm}). The coefficients corresponding to the reflections  from panels B, C and D (see Fig.~\ref{FIG2}) are expressed in terms of the coefficients for the process A which are given explicitly by Eq.~(\ref{coeff}).}
\label{TBL1}
\end{table}

Next we introduce two pairs of symmetrized states defined as
\beqa
\Psi_{\bf k}^{1\pm}({\bf r}) = \frac{1}{\sqrt{1+a^2}}\left[\left(1-\frac{ab}{\sqrt{1-b^2}}\right)~\psi^{AC\pm} + \frac{a}{\sqrt{1-b^2}}~\psi^{BD\mp}\right],  \nonumber \\
\Psi_{\bf k}^{2\pm}({\bf r}) =
\frac{1}{\sqrt{1+a^2}}\left[\left(a+\frac{b}{\sqrt{1-b^2}}\right)~\psi^{AC\pm}
- \frac{1}{\sqrt{1-b^2}}~\psi^{BD\mp}\right],          \label{psisyp}
\eeqa
where $\psi^{AC\pm}= 1/\sqrt{N}(\psi^A\pm\psi^C)$ and
$\psi^{BD\mp}=1/\sqrt{N^{\prime}}(\psi^B\mp\psi^D)$, with N and
$N^{\prime}$ being normalization factors that insure
$\langle\psi^{\alpha}|\psi^{\alpha}\rangle = 1$ for both linear
combinations. The parameter 'b' is set equal to the  scalar product
of the two sets of linear combinations, $b=
\langle\psi^{AC\pm}|\psi^{BD\mp}\rangle$, insuring the
orthonormality of
the eigenstates,
\beq
\langle\Psi^{i+}|\Psi^{j+}\rangle = \delta_{ij}, ~~~~~~~~~~~~~~\langle\Psi^{i-}|\Psi^{j-}\rangle = \delta_{ij},
\eeq where $\delta_{ij}$ is the Kronecker
$\delta$-symbol. By varying the parameter 'a' we can change the
weight of the heavy-hole and light-hole modes that contribute to a
particular eigenfunction. For a state given by Eq.~(\ref{psisymm})
the weight of the heavy-holes is defined as \beq W_{HH}^{\alpha} =
|c_1^{\alpha}|^2 + |c_2^{\alpha}|^2 + |c_5^{\alpha}|^2 +
|c_6^{\alpha}|^2,  \label{whh} \eeq while for the light-holes we
have $W_{LH}^{\alpha} = 1-W_{HH}^{\alpha}$. We choose the parameter
'a'  to maximize the weight of the heavy-holes in the eigenstates
$\Psi^{1\pm}$ and the weight of the light-holes in $\Psi^{2\pm}$. The
convenience of this choice will become clear once we address the
problem of counting the eigenstates. The explicit value of the
parameter 'a' satisfying this condition is \beq a = \frac{1}{2
W^{\prime}}\left(W_{HH}^{BD\mp} - W_{HH}^{AC\pm} + \sqrt{(W_{HH}^{BD\mp} -
W_{HH}^{AC\pm})^2 + 4(W^{\prime})^2}\right),
\eeq
where $W^{\prime} =
c_1^{AC\pm}c_1^{BD\mp} + c_2^{AC\pm}c_2^{BD\mp} +c_5^{AC\pm}c_5^{BD\mp} +
c_6^{AC\pm}c_6^{BD\mp}$. Notice that, while in general the
eigenstates
$\Psi^{i\pm}$ represent a superposition of both heavy-hole and
light-hole modes, in the particular case of normal incidence,
$\theta=0$, $\Psi^{1\pm}$ contains only heavy-holes and $\Psi^{2\pm}$
contains only light-holes. In addition, for $\theta=\theta_c$,
$\Psi^{2\pm}$ reduces again to a superposition of light-holes
propagating parallel to the boundary, {\em i.e.}, $\phi(\theta_c) =
\pi/2$.
There is a symmetry  between the modes contributing to $\Psi^{i+}$
 and the states with opposite helicity contributing to
$\Psi^{i-}$. This symmetry  translates into a set of
relation between the corresponding coefficients that is
summarized in Table \ref{TBL2}.
\begin{table}[h!t!]
\begin{tabular}{|c||c|c|c|c||c|c|c|c|}
\hline
~ & $c_1^{\alpha}$ & $c_2^{\alpha}$ & $c_3^{\alpha}$ & $c_4^{\alpha}$ & $c_5^{\alpha}$ & $c_6^{\alpha}$ & $c_7^{\alpha}$ & $c_8^{\alpha}$ \\
\hline \hline
$\Psi^{1+}$ & $c_1^1$ & $c_2^1$ & $c_3^1$ &  $c_4^1$ &  $c_1^1$ & $-c_2^1$ & $-c_3^1$ &  $c_4^1$ \\
\hline
$\Psi^{1-}$ & $-c_2^1$ & $c_1^1$ & $c_4^1$ &  $-c_3^1$ &  $c_2^1$ & $c_1^1$ & $c_4^1$ &  $c_3^1$ \\
\hline
$\Psi^{2+}$  & $c_1^2$ & $c_2^2$ & $c_3^2$ &  $c_4^2$ &  $c_1^2$ & $-c_2^2$ & $-c_3^2$ &  $c_4^2$ \\
\hline
$\Psi^{2-}$  & $-c_2^2$ & $c_1^2$ & $c_4^2$ &  $-c_3^2$ &  $c_2^2$ & $c_1^2$ & $c_4^2$ &  $c_3^2$ \\
\hline
\end{tabular}
 \caption{Relations between the coefficients of the symmetric eigenstates for an incident angle for the heavy-holes $\theta < \theta_c$. The wave functions for the pairs of states $\Psi_k^{1(2)+}$   and $\Psi_k^{1(2)-}$ (see Eq.~(\ref{psisyp})) are written in the general form given by Eq.~(\ref{psisymm}).
The existence of two pairs of  modes labeled by 1 and 2 is reminiscent of the heavy- and light-holes of the bulk, while the degree of freedom labeled by '+' and '-' is the analogous of helicity.
Notice that all the coefficients can be expressed in terms of two sets of four parameters, one set for each pair of modes $\alpha =1\pm$ and $\alpha =2\pm$, respectively.}
\label{TBL2}
\end{table}

So far we have addressed the problem of finding a set of  symmetric
eigenstates in  the case when the incident angle for the
heavy-holes is smaller than the critical angle, $\theta < \theta_c$.
 Let us focus now on the case $\theta > \theta_c$, when the
light-holes are localized near the boundary. There are two main
differences between the two cases. The first one concerns the number
of linearly independent eigenstates. Far from the boundary, the
system should be locally equivalent to an infinite system and,
consequently, the number of degrees of freedom should be the same.
The set of four independent eigenstates that we obtained for $\theta
< \theta_c$ corresponds to the double degenerate heavy-holes and
light-holes of the infinite system. In contrast, for $\theta >
\theta_c$ only the heavy-holes modes propagate in the z-direction
and, therefore, we expect to have only two independent eigenfunction
for a given incident angle. The second difference is a formal one
and consist in the spinors $V_i(\chi)$ being complex, in contrast
with $U_{\lambda}$ which are real. Consequently, the coefficients of
the eigenstates involving localized modes will be complex numbers.
The coefficients $A_i$, $B_i$ for the wave function (\ref{psihp2})
are given explicitly in appendix \ref{AppB}. The symmetric
eigenstates can be written again as superposition of reflection
processes analogous to those in Fig.~\ref{FIG2}, except that this time
the light-holes are localized modes confined near the boundary. The
most general form of such an eigenstate can be written as \beqa
\Psi_{\bf k}^{\alpha}({\bf r}) = \frac{1}{\sqrt{\Omega}}e^{ik_x x}\left\{\left[c_1^{\alpha}~U_{H+}(\theta)+ c_2^{\alpha}~U_{H-}(\theta)\right]e^{ik_z z}\right. &+& \left[c_3^{\alpha}~V_1(\chi)+ c_4^{\alpha}~V_2(\chi)\right]e^{-Q z}  \label{psisymm2} \\
&+& \left.\left[c_5^{\alpha}~U_{H+}(\pi-\theta) +
c_6^{\alpha}~U_{H-}(\pi-\theta)\right]e^{-ik_z z}\right\}, \nonumber
\eeqa where $Q(k, \theta)$ and $\chi(\theta)$ are given by
Eq.~(\ref{pphi2}). Again, the coefficients corresponding to a family
of reflection processes similar to that represented in Fig.~\ref{FIG2}
are not all independent and we summarize the relations between them
in Table \ref{TBL3}.
\begin{table}[h!b!p!]
\begin{tabular}{|c||c|c||c|c||c|c|}
\hline
~ & $c_1^{\alpha}$ & $c_2^{\alpha}$ & $c_3^{\alpha}$ & $c_4^{\alpha}$ & $c_5^{\alpha}$ & $c_6^{\alpha}$ \\
\hline \hline
$\psi^A$ & 1 & 0 & $B_1$ & $B_2$ & $A_1$ & $A_2$  \\
\hline
$\psi^B$ & 0 & 1 & i$B_1$ & -i$B_2$ & $-A_2$ & $A_1$ \\
\hline
$\psi^C$ & $A_1^*$ & $-A_2^*$ & i$B_1^*$ & i$B_2^*$ & 1 & 0 \\
\hline
$\psi^D$ & $A_2^*$ & $A_1^*$ & -$B_1^*$ & $B_2^*$ & 0 & 1  \\
\hline
\end{tabular}
\caption{Coefficients for a  family of wave-functions with localized
light-hole modes analogous to the scattering processes represented
in Fig.~\ref{FIG2}. The wave functions are expressed in the general
form given by Eq.~(\ref{psisymm2}). The complex coefficients
corresponding to the reflection of a heavy-hole with helicity -3/2
(analog to the process represented in Fig.~\ref{FIG2} panel B), as
well as the ``re-combination'' processes (see Fig.~\ref{FIG2} panels
C and D), are expressed in terms of the coefficients for the
heavy-hole +3/2 reflection (panel A) given by Eq.~(\ref{coeff2}). }
\label{TBL3}
\end{table}

Finally, we define the symmetryzed eigenstates
\beqa
\Psi_{\bf k}^1({\bf r}) &=& \frac{1}{\sqrt{C_1}}\left[\psi^{AC+} + i\psi^{BD+}\right], \nonumber \\
\Psi_{\bf k}^2({\bf r}) &=& \frac{1}{\sqrt{C_2}}\left[\psi^{AC+} - i\psi^{BD+}\right],     \label{psiloc}
\eeqa
where $C_1$ and $C_2$ are normalization constants and the linear
combinations $\psi^{AC+} = \psi^A + \psi^C$ and
$\psi^{BD+} = \psi^B + \psi^D$ contain wave-functions with
coefficients given in Table \ref{TBL3}. The wave-functions
 defined by Eq.~(\ref{psiloc}) represent a complete system of
eigenfunctions in the  subspace characterized by $k=|{\bf k}|$ and
the  incident angle $\theta>\theta_c$. The symmetry between
heavy-hole contributions with opposite helicity is reflected by the
special relation between the corresponding coefficients,
$c_2 = \pm i~c_1$ and $c_6 = \pm i~c_5$. The
complete set of relations between the coefficients is given in Table
\ref{TBL4}.
\begin{table}[h!t!]
\begin{tabular}{|c||c|c||c|c||c|c|}
\hline
~ & $c_1^{\alpha}$ & $c_2^{\alpha}$ & $c_3^{\alpha}$ & $c_4^{\alpha}$ & $c_5^{\alpha}$ & $c_6^{\alpha}$ \\
\hline \hline
$\Psi^1$ & $c_1^1$ & i$c_1^1$ & 0 & (1+i)$b_2^1$ & $(c_1^1)^*$ & i$(c_1^1)^*$  \\
\hline
$\Psi^2$  & $c_1^2$ & -i$c_1^2$ & (1+i)$b_1^2$ & 0 & $(c_1^*)^*$ & -i$(c_1^2)^*$  \\
\hline
\end{tabular}
\caption{Relations between the coefficients of the symmetric
eigenstates in the presence of the localized modes, i.e. for an
incident angle for the heavy-holes $\theta>\theta_c$. The wave
functions for the eigenstates $\Psi^1$ and $\Psi^2$ given by Eq.
(\ref{psiloc})are expressed in
the general form given by Eq.~ (\ref{psisymm2}). The coefficients
$b_2^1$ and $b_1^2$ are real. Notice that there are only two
independent eigenstates for a given set of parameters (k, $\theta$),
in contrast with the case $\theta<\theta_c$ when four independent
eigenvectors were found.} \label{TBL4}
\end{table}

We have identified complete sets of linearly independent
eigenstates for arbitrary values of the incident momentum of the
heavy-holes, i.e. for arbitrary parameters (k, $\theta$). These
sets of eigenstates are given by the equations (\ref{psisyp}) and
(\ref{psiloc}). In order
to construct an orthonormal  basis for the quantum problem defined
by the Luttinger Hamiltonian in half-space, we have to solve two
problems: a) the orthogonality of the eigenstates, and b) the proper
counting of the modes. We address these issues in the remaining of
this section.

\subsection{Orthogonality of the symmetrized eigenstates}
\label{Sec:Orthogonality}

Let us start by clarifying a few aspects related to the definition
of the scalar product that we use here. Consider two wave-functions
$\Psi_{\bf k}^{\alpha}$ and $\Psi_{\bf
k^{\prime}}^{\alpha^{\prime}}$ where ${\bf k} = (k_x, 0 ,k_z)$, with
$k_x/k_z=-\tan(\theta)$, represents the momentum of the incident
heavy-hole mode and $\alpha\in\{1+, 1-, 2+, 2-\}$, if $\theta
<\theta_c$ or $\alpha\in\{1, 2\}$, if $\theta >\theta_c$. For
reasons that will become clear below when we address the problem of
counting the states, we consider a ``discretized'' momentum space so
that two momenta are considered  identical, ${\bf k} = {\bf
k^{\prime}}$,
 if they are both in a certain cell of volume $\delta k^3$ and
different otherwise. We define the scalar product of the two states
as
\beq \langle \Psi_{\bf k}^{\alpha} | \Psi_{\bf
k^{\prime}}^{\alpha^{\prime}} \rangle = \int_{\Omega}d^3r~ \left[
\Psi_{\bf k}^{\alpha}({\bf r})\right]^{\dagger} \Psi_{\bf
k^{\prime}}^{\alpha^{\prime}}({\bf r}), \label{scalprod}
\eeq
where
$\Omega=L^3$ is the volume of the system. We will always consider
the thermodynamic limit $L\rightarrow \infty$, so that
$L\delta k \gg 1$. Due to the particular form of the eigenfunctions
(\ref{psisymm}) and (\ref{psisymm2}), the integrations over x, y,
and z can be performed separately. Due to our choice for the
coordinate system,
the wave-functions are independent of y and therefore the
integration over this coordinate trivially generates a factor L.
Further, all the x-dependence is contained in the factors $\exp(i
k_x x)$ so that  the x-integration generates a factor L$\delta_{k_x
k_x^{\prime}}$. Consequently, Eq.~(\ref{scalprod}) reduces to the
one dimensional integral \beq \langle \Psi_{\bf k}^{\alpha} |
\Psi_{\bf k^{\prime}}^{\alpha^{\prime}} \rangle =
L^2\delta_{k_x,k_x^{\prime}} \int_0^L dz~ \left[ \Psi_{\bf
k}^{\alpha}(z)\right]^{\dagger} \Psi_{\bf
k^{\prime}}^{\alpha^{\prime}}(z) , \label{scalprod1} \eeq where
$\Psi_{\bf k}^{\alpha}(z) = \Psi_{\bf k}^{\alpha}({\bf r})\exp(-ik_x
x)$. We notice that, in order to insure convergence, all the
z-dependent oscillatory terms in Eq.~(\ref{scalprod1}) are
multiplied by a convergence factor $\exp(-\eta z)$ with $\eta L
\gg 1$, and the limit $\eta \rightarrow 0$ is taken at the end of
the calculation. Considering  now  the expressions (\ref{psisymm})
and (\ref{psisymm2}) for the eigenfunctions, together with the
properties of the coefficients summarized in Table \ref{TBL2} and
Table \ref{TBL4}, as well as the expressions for the spinors given
in Appendix \ref{AppA}, we obtain the generic expressions of the terms
contributing to the scalar product ({\ref{scalprod1}). We will
concentrate on the case $k_x=k_x^{\prime}$, because for different
x-components of the momentum the states are manifestly orthogonal.
Also notice that, due to the exponential factors $\exp(-Q z)$, the
localized modes from the eigenfunctions (\ref{psisymm2}) will
generate terms proportional to $1/(LQ)$, which vanish in the
thermodynamic limit, and therefore will not contribute to the scalar
products. Omitting these terms, the generic contribution to the
integral over z in the scalar product has the form
\beq L^2 \left[
\Psi_{\bf k}^{\alpha}(z)\right]^{\dagger} \Psi_{\bf
k^{\prime}}^{\alpha^{\prime}}(z) = \sum_{(k_1, k_2)}\frac{1}{L}\left[
\Lambda_1^{\alpha, \alpha^{\prime}}(k_1, k_2)\cos[(k_1-k_2)z] +
\Lambda_2^{\alpha, \alpha^{\prime}}(k_1, k_2)\sin[(k_1-k_2)z]
\right],   \label{scalprod2}
\eeq
with $(k_1, k_2) \in \{(k_z, \pm
k_z^{\prime}), (k_z, \pm q_z^{\prime}), (q_z, \pm k_z^{\prime}),
(q_z, \pm q_z^{\prime})\}$, where $k_z$ and $k_z^{\prime}$ are the
z-components of the incident wave-vector for the heavy-hole modes,
while $q_z$ and
$q_z^{\prime}$ are the z-components of the incident light-hole waves.
The reflected waves  have components with opposite signs.
Due to the symmetry properties of the
eigenstates, some of the coefficients $\Lambda_i^{\alpha,
\alpha^{\prime}}$ are identically zero.  The non-vanishing
coefficients are summarized in Table \ref{TBL5}.
\begin{table}[h!t!]
\begin{tabular}{|c||c|c|c|c|c|c|}
\hline
~ & (1+) & (1-) & (2+) & (2-) & (1) & (2)  \\
\hline \hline
(1+) & $\Lambda_1$ &  $\Lambda_2$ & $\Lambda_1$ & $\Lambda_2$ & $\Lambda_1$, $\Lambda_2$ & $\Lambda_1$, $\Lambda_2$   \\
\hline
(1-) & $\Lambda_2$ &  $\Lambda_1$ & $\Lambda_2$ & $\Lambda_1$ & $\Lambda_1$, $\Lambda_2$ & $\Lambda_1$, $\Lambda_2$   \\
\hline
(2+) & $\Lambda_1$ &  $\Lambda_2$ & $\Lambda_1$ & $\Lambda_2$ & $\Lambda_1$, $\Lambda_2$ & $\Lambda_1$, $\Lambda_2$   \\
\hline
(2-) & $\Lambda_2$ &  $\Lambda_1$ & $\Lambda_2$ & $\Lambda_1$ & $\Lambda_1$, $\Lambda_2$ & $\Lambda_1$, $\Lambda_2$   \\
\hline
(1) & $\Lambda_1$, $\Lambda_2$ &  $\Lambda_1$, $\Lambda_2$ & $\Lambda_1$, $\Lambda_2$ & $\Lambda_1$, $\Lambda_2$ & $\Lambda_1$, $\Lambda_2$ & 0   \\
\hline
(2) & $\Lambda_1$, $\Lambda_2$ &  $\Lambda_1$, $\Lambda_2$ & $\Lambda_1$, $\Lambda_2$ & $\Lambda_1$, $\Lambda_2$ & 0 & $\Lambda_1$, $\Lambda_2$  \\
\hline
\end{tabular}
 \caption{Non-vanishing $\Lambda_i^{\alpha, \alpha^{\prime}}(k_1, k_2)$ coefficients that contribute to the scalar product of the  eigenstates $\Psi_{\bf k}^{\alpha}(z)$ and $\Psi_{\bf k}^{\alpha^{\prime}}(z)$.  The rows of the table are indexed by $\alpha$ and the columns by $\alpha^{\prime}$. For example, the scalar product of $\Psi^{2+}$ and $\Psi^{2-}$ will be an integral over z of a sum containing only $|lambda_2$-type terms. The simple z-dependence of the integrand, given by Eq.~(\ref{scalprod2}), and the properties of the coefficients $\Lambda_i^{\alpha, \alpha^{\prime}}(k_1, k_2)$ enable us to prove the orthogonality of the eigenfunctions.}
\label{TBL5}
\end{table}
Integrating now Eq. (\ref{scalprod2}) over z, and
taking the necessary limit for the
convergence factor $\eta$, we obtain two types of
contributions to  the scalar product of two eigenstates
$\Psi_{\bf k}^{\alpha}$ and
$\Psi_{{\bf k}^{\prime}}^{\alpha^{\prime}}$:
$\Lambda_1^{\alpha,\alpha^{\prime}}(k_1, k_1)$ if $k_1=k_2$ and
$\frac{1}{L}~\Lambda_2^{\alpha,
\alpha^{\prime}}(k_1, k_2)/(k_1-k_2)$ if $k_1\neq k_2$. For the
contribution proportional to $\Lambda_2$  we distinguish two regimes.
 The first regime  is
characterized by a finite difference $k_1-k_2$,  so that
$L|k_1-k_2| \gg 1$, and occurs for example in
combinations like $k_z+k_z^{\prime}$ or $k_z+k_z^{\prime}$. In this
case  there is no contribution to the scalar product in the
thermodynamic limit.
In the second regime, the difference between the two components of
the wave-vector can be arbitrarily small, so that
 $k_1\approx k_2$ and the ratio $\Lambda_2^{\alpha,
\alpha^{\prime}}(k_1, k_2)/(k_1-k_2)$ seems to diverge. However,
by evaluating explicitly the coefficients in this limit we obtain
 $\Lambda_2^{\alpha, \alpha^{\prime}}(k_1, k_2)
\propto k_1-k_2$ and the corresponding contribution to the scalar
product vanishes again in the thermodynamic limit. This behavior
is valid for all cases, except when
$(\alpha,\alpha^{\prime}) = (i\pm, j\mp)$ with $i\neq j$, for
example in the case $k_z\approx k_z^{\prime}$, $q_z\approx q_z^{\prime}$ and $(\alpha,\alpha^{\prime}) = (1+, 2-)$.
In these special
cases when the $\Lambda_2$ terms do not vanish individually, a
direct evaluation of the coefficients in the limit
 $k_z\rightarrow k_z^{\prime}$ shows that
\beq
\frac{\Lambda_2^{\alpha,\alpha^{\prime}}(q_z,
q_z^{\prime})}{\Lambda_2^{\alpha,\alpha^{\prime}}(k_z,
k_z^{\prime})} = -\left|\frac{\partial q_z}{\partial k_z}\right| =
-\frac{\xi |\cos(\theta)|}{\sqrt{\xi-\sin^2(\theta)}}. \eeq
Consequently,
the total contribution to the scalar product,
$\frac{1}{L}[\Lambda_2^{\alpha,\alpha^{\prime}}(k_z,
k_z^{\prime})/(k_z-k_z^{\prime}) +
\Lambda_2^{\alpha,\alpha^{\prime}}(q_z,
q_z^{\prime})/(q_z-q_q^{\prime})]$ vanishes again in the
thermodynamic limit. The $\Lambda_1^{\alpha, \alpha^{\prime}}(k_1,
k_1)$ terms also vanish for $\alpha \neq \alpha^{\prime}$. In fact,
the only contributions to the scalar products that survive in the
thermodynamic limit arise from $\Lambda_1^{\alpha, \alpha}(k_z,
k_z)$ and $\Lambda_1^{\alpha, \alpha}(q_z, q_z)$ and we have \beqa
\Lambda_1^{\alpha, \alpha}(k_z, k_z) &+& \Lambda_1^{\alpha, \alpha}(q_z, q_z) = \sum_{i=1}^8~\left| c_i^{\alpha}  \right|^2 = 1~~~~~~~~~\mbox{for}~~~\theta<\theta_c,  \nonumber \\
\Lambda_1^{\alpha, \alpha}(k_z, k_z) &=& \left| c_1^{\alpha}
\right|^2 + \left| c_2^{\alpha}  \right|^2 + \left| c_5^{\alpha}
\right|^2 + \left| c_6^{\alpha}  \right|^2 = 1
~~~~~~\mbox{for}~~~\theta>\theta_c, \eeqa where the unity was
obtained by a proper choice of the normalization constants for the
coefficients $c_i^{\alpha}$ in the construction of the symmetrized
states. We conclude that the symmetrized
eigenstates constructed here represent an orthonormal system
satisfying $\langle \Psi_{\bf k}^{\alpha} | \Psi_{\bf
k^{\prime}}^{\alpha^{\prime}} \rangle = \delta_{\alpha
\alpha^{\prime}}\delta_{\bf k k^{\prime}}$.

\subsection{Counting the eigenstates}
\label{Sec:Counting}

The standard procedure for counting the states and transforming sums
over the k-vector into integrals for a quantum system  in the
thermodynamic limit consists in imposing periodic boundary
conditions for a  finite volume $\Omega = L^3$ system, and then
taking the limit $L\rightarrow \infty$. The periodic boundary
conditions generate a discrete set of wave-vectors ${\bf k} = (k_x,
k_y, k_z) = 2\pi/L(n_1, n_2, n_3)$, where $n_i$ are integers. The
procedure works in the presence of spin-orbit coupling if no
boundary is considered and the eigenstates are the heavy-hole and
light-hole modes. However in our case, the eigenstates are certain
combinations of heavy- and light-holes  corresponding to the same
energy. Explicitly, for $\theta<\theta_c$, heavy-hole modes with
${\bf k}_{H}=2\pi/L(n_x, n_y, n_z)$ have to be paired up with
light-holes characterized by the wave-vector ${\bf
q}_{L}=2\pi/L(n_x, n_y, m_z)$ with $m_z$ satisfying \beq
\xi\left(\frac{2\pi}{L}\right)^2\left(n_x^2+n_y^2+ n_z^2\right) =
\left(\frac{2\pi}{L}\right)^2\left(n_x^2+n_y^2+ m_z^2\right). \eeq
For an arbitrary value of $\xi=m_L/m_H$ there is no integer solution
of this equation. To overcome this difficulty we consider a
partition of the  reciprocal space with the property that all the
heavy-hole modes characterized by helicity $\lambda$ and
wave-vectors situated in a cell $\delta k^3$ centered on ${\bf k}$
are represented by a single heavy-hole mode
 $({\bf k}, \lambda)$ with the energy $\epsilon_k = k^2/2 m_H$ and an
effective ``degeneracy'' $\nu_0 = \delta k^3 (L/2\pi)^3 \gg 1$. In the
thermodynamic limit we can choose $\delta k$ arbitrarily small, so
that $1/\delta k$ is larger than any relevant length-scale in the
problem. For incident angles $\theta<\theta_c$, the heavy-hole modes
from a cell $({\bf k}, \delta k^3)$ will pair with light-hole modes
from a cell $({\bf q}, \delta q^3)$, were all the points of the new
cell were obtained by mapping the points of the original cell, ${\bf
k}\rightarrow {\bf q}$, using Eq.~ (\ref{pphi2}). The ratio of the
light-hole and heavy-hole degrees of freedom from the two cells is
\beq R_{LH} =\frac{\delta q^3}{\delta k^3}=\left| \frac{\partial
q_z}{\partial k_z} \right| = \frac{\xi
|\cos(\theta)|}{\sqrt{\xi-\sin^2(\theta)}}. \eeq For each heavy-hole
mode with helicity $+3/2$
used in the construction of the eigenstates (\ref{psisyp}) we
will use another heavy-hole with opposite helicity as well as
$2R_{LH}$ light-hole modes and we will generate $2\mu_1$ states
$\Psi_k^{1\pm}$ and $2\mu_2$ states $\Psi_k^{2\pm}$. The effective
degeneracies $\mu_i$ are determined by the condition that, far from
the boundary, the number of heavy- and light-hole modes is the same
as in an infinite system. Consequently we have \beqa
\mu_1 W_{HH}^{1\pm} + \mu_2 W_{HH}^{2\pm} &=& 1, \nonumber \\
\mu_1 W_{LH}^{1\pm} + \mu_2 W_{LH}^{2\pm} &=& R_{LH},
\eeqa
where $ W_{LH}$ and $ W_{HH}$ are the light-hole and heavy-hole
weights, respectively. We obtain for the effective degeneracies
of the eigenstates $\Psi_{\bf k}^{i\sigma}$ given by (\ref{psisyp})
 the expressions
\beq
\mu_1 =\frac{1-W_{HH}^{2\sigma}(1+R_{LH})}{W_{HH}^{1\sigma}-W_{HH}^{2\sigma}}, ~~~~~~~~~~~~~~~~~~~~~\mu_2 = 1+R_{LH}-\mu_1. \label{mueff}
\eeq
Notice that these expressions are meaningful only if the effective
degeneracies satisfy the condition $0 \leq \mu_i \leq (1+R_{LH})$.
Consequently, the weights of the heavy-hole modes contained in
eigenstates defined by equations (\ref{psisyp})
 cannot be arbitrary. If, for example,
$W_{HH}^{1\sigma} > W_{HH}^{2\sigma}$ they have to satisfy the inequalities
\beq
W_{HH}^{1\sigma} > \frac{1}{1+R_{LH}}, ~~~~~~~~~~~~~~~~~~~~~~
W_{HH}^{2\sigma} < \frac{1}{1+R_{LH}}.         \label{ineq}
\eeq
These inequalities are  the reason behind our choice for the
parameter 'a' in
the construction of the eigenstates (\ref{psisyp}). By maximizing
the weight of the heavy-holes in the type-1 eigenstates and
minimizing it in the type-2 eigenstates we insure that the
inequalities (\ref{ineq}) are always satisfied.
The analysis of the  case $\theta > \theta_c$ is straightforward,
because far from the boundary only the heavy-hole modes survive.
Consequently, there is  a direct correspondence between the
number of degrees of freedom associated with the
heavy-holes and the number of symmetrized eigenstates (\ref{psiloc}).

In this section we have constructed an orthonormal basis for
the Luttinger Hamiltonian in half space. For our particular
choice of coordinates, the basis consists in the symmetrized
eigenfunctions $\Psi_{\bf k}^{i\sigma}$ given by equations
(\ref{psisyp})  with $i=1,2$, $\sigma=\pm$, and
${\bf k} = (k_x, 0, k_z)$ for incident angles
 $\theta = -\arccos(k_z/k) < \theta_c$,
together with the eigenstates $\Psi_{\bf k}^{i}$ given by Eq.
(\ref{psiloc}) with $i=1,2$ and incident angles
$\theta = -\arccos(k_z/k) > \theta_c$. The eigenstates from the
first set   have an effective degeneracy $\nu_0 \mu_i$ with
$\mu_i$ given by (\ref{mueff}), while the eigenstates from the
second set have an effective degeneracy $\nu_0$ in our
discretization  construction for the momentum space.
 Notice that the overall factor
 $\nu_0$ is irrelevant in the calculation of the physical
quantities. We conclude  this section by giving the rules for
transforming sums over the k-vector into integrals. Let us assume
that f({\bf k}, $\alpha$) is a function that depends on the diagonal
matrix elements $\langle \Psi_{\bf k}^{\alpha}|A|\Psi_{\bf
k}^{\alpha}\rangle$ of a certain  operator A and that we are
interested in calculating  its sum over all the eigenstates of
the basis. We have \beq \sum_{\bf k}^{(\theta < \theta_c)}\sum_{i,
\sigma}(\nu_0\mu_i) f({\bf k}, i, \sigma) ~~+ \sum_{\bf k}^{(\theta
> \theta_c)}\sum_{i}(\nu_0) f({\bf k}, i) =
\frac{\Omega}{(2\pi)^3}\sum_{i, \sigma}\int_{(\theta <
\theta_c)}d^3k ~\mu_i f({\bf k}, i, \sigma) +
\frac{\Omega}{(2\pi)^3}\sum_{i}\int_{(\theta > \theta_c)}d^3k
~f({\bf k}, i), \label{sumint} \eeq where we have taken into account
that the volume of a cell in the discretized momentum space is
$\delta k^3 = \nu_0(2\pi/L)^3$ and we
have considered the thermodynamic limit $L\rightarrow\infty$. For
example if $f({\bf k}, \alpha)$ represents the occupation number at
zero temperature of the mode with the quantum numbers $({\bf k},
\alpha)$, by applying Eq. (\ref{sumint})we can easily obtain the
relation between the Fermi
k-vector and the density of particles, $n = k_F^3(1 +
\xi^{\frac{3}{2}})/(6\pi^2)$. Notice that each component $\Psi_{\bf
k}^{\alpha}$ of the basis, although contains a superposition of
heavy- and light-holes, is labeled by the k-vector of the incident
heavy-hole mode and has an energy $\epsilon_k = k^2/(2m_H)$
independent of $\alpha$. Consequently, the Fermi wave-vector $k_F$
is independent of $\alpha$ and is determined by the density of
particles and the spin-orbit coupling.

\section{Oscillating spin density and spin accumulation}
\label{Sec:Many-body}

In this section we investigate the properties of the spin density
and explore the possibility of spin accumulation in a system with a
sharp boundary described by the Luttinger model (here and below we
actually study the accumulation of the total orbital momentum, but
we call it ``spin accumulation'' for brevity). We can derive the
spin density from the one-particle propagators. Using the
orthonormal basis constructed in the previous section, we can define
the Green's functions for the Luttinger model in half space. The
retarded Green's function is \beq G_{ab}^{(R)}(\omega;~{\bf r},{\bf
r}^{\prime}) = \sum_{{\bf k}, \alpha} \nu_{k
\alpha}\frac{\left[\Psi_{\bf k}^{\alpha}({\bf
r}^{\prime})\right]_b^*\left[\Psi_{\bf k}^{\alpha}({\bf
r})\right]_a}{\omega -\mu -\epsilon_k+i 0^+}, \eeq where $\mu$ is
the chemical potential and
 $\Psi_{\bf k}^{\alpha}({\bf r})$ is a vector from the basis
defined by the  equations (\ref{psisyp})
and (\ref{psisymm2})
corresponding to the eigenvalue $\epsilon_k = k^2/2m_H$, with $m_H$
being the heavy-hole mass. The eigenvectors are labeled by the
wave-vector ${\bf k}$ of the incident heavy-hole mode and the index
$\alpha$ which takes four  values, $\alpha \in \{1+, ~1-, ~2+,
~2-\}$, if all the modes  propagate in the z-direction ($\theta <
\theta_c$), and two values, $\alpha \in \{1, ~2\}$, if the
eigenstate contains localized modes ($\theta > \theta_c$). In
addition, because the states $\Psi_{\bf k}^{\alpha}({\bf r})$ are
four-component spinors, we have the indices $a$ and $b$ to label
their components. Finally, to correctly account  for the light- and
heavy-hole modes used in the construction of the basis, we have the
effective degeneracy factor $\nu_{k \alpha} = \nu_0\mu_{\alpha}$,
where $\mu_{\alpha}$ is given by Eq.~(\ref{mueff}) when the
eigenvector contains only propagating modes, i.e. for
$\theta<\theta_c$, and $\mu_{\alpha}=1$ otherwise, while $\nu_0 =
\delta k^3/(2\pi/L)^3$ is determined by our momentum discretization
procedure (as we have noted above this factor drops out of all final
results).
 The density of the total momentum S can be expressed in terms of
Green's function as $\langle {\bf S}\rangle({\bf r}) = -i\int
d\omega\mbox{Tr}[\hat{\bf S}\hat{G}(\omega; {\bf r, r})]$, where the
matrices ${\bf S}_{ab}= ([S_x]_{ab},  [S_y]_{ab}, [S_z]_{ab})$ are
given in Appendix \ref{AppA}. In terms of the eigenstates for the
Luttinger model in half-space we have explicitly \beq \langle
S_l\rangle({\bf r}) = \sum_{\bf k}\sum_{\alpha} \nu_{k \alpha}
\left(\left[\Psi_{\bf k}^{\alpha}({\bf r})\right]^{\dagger}
\hat{S}_l \Psi_{\bf k}^{\alpha}({\bf r})\right) n_{\alpha}({\bf k}),
\label{jdens} \eeq where $\hat{S}_l$ is the matrix for the $l\in\{x,
y, z\}$ component of the total momentum, $n_{\alpha}({\bf k})\in
\{0, 1\}$ is the zero temperature occupation number of the
$\Psi_{\bf k}^{\alpha}$ state with effective degeneracy $\nu_{k
\alpha}$, and the summations, which include all the eigenstates in
the basis, can be converted into integrals using Eq.~(\ref{sumint}).
The analysis can be greatly simplified if we take into account the
symmetry properties of the eigenstates. These symmetries translate
directly into relations between single state contributions to the
matrix element of the spin density, \begin{equation}
\label{matrixel} {\cal S}_i^{\alpha}({\bf k, r}) = \left[\Psi_{\bf
k}^{\alpha}({\bf r})\right]^{\dagger} \hat S_i \Psi_{\bf
k}^{\alpha}({\bf r}).
\end{equation}
Explicitly we have \beqa &~&\left\{
\begin{array}{ll}
{\cal S}_x^{i~\sigma}({\bf k, r}) = -{\cal S}_x^{i~(-\sigma)}({\bf k, r}) ~~~~~~~~~~~~ \mbox{for}~~\theta<\theta_c, \\
{\cal S}_x^{i~~}({\bf k, r}) = 0 ~~~~~~~~~~~~~~~~~~~~~~~~~~~~~~ \mbox{for}~~\theta>\theta_c,
\end{array}\right. \nonumber \\
&~&\left\{ \begin{array}{ll}
{\cal S}_z^{i~\sigma}({\bf k, r}) = 0 ~~~~~~~~~~~~~~~~~~~~~~~~~~~~~~ \mbox{for}~~\theta<\theta_c, \\
{\cal S}_z^{i~~}({\bf k, r}) = 0 ~~~~~~~~~~~~~~~~~~~~~~~~~~~~~~ \mbox{for}~~\theta>\theta_c,
\end{array}\right.
\eeqa
where $i = 1,2$ and $\sigma = \pm$ represent the labels of the
corresponding eigenstates. We conclude that the contribution
to the x and z-components of the spin density from any wave-vector
${\bf k}$ is identically zero. On the other hand, for the
y-component we obtain
\beq
\left\{ \begin{array}{ll}
{\cal S}_y^{i~\sigma}({\bf k, r}) = {\cal S}_y^{i~(-\sigma)}({\bf k, r}) ~~~~~~~~~~~~~~ \mbox{for}~~\theta<\theta_c, \\
{\cal S}_y^{1~~}({\bf k, r}) = {\cal S}_y^{2~~}({\bf k, r}) ~~~~~~~~~~~~~~~~~~ \mbox{for}~~\theta>\theta_c,
\end{array}\right.
\eeq and consequently we get in general a non-zero total
contribution for a given ${\bf k}$. In other words, in a system
described by the Luttinger Hamiltonian and having a sharp planar
boundary, the contribution to the spin density vector from states
characterized by a wave-vector ${\bf k}$ is oriented along a
direction parallel to the boundary and perpendicular to the
wave-vector, $\vec{\cal S}({\bf k, r}) \propto {\bf n}_b\times {\bf
k}$, where ${\bf n}_b$ is a unit vector normal to the boundary. We
remind here that the eigenfunctions $\Psi_{\bf k}^{\alpha}$ are
written in the ``local''  coordinate system determined by the
convention ${\bf k} = (k_x>0, 0, k_z)$, and therefore, whenever we
have to use a system of coordinates independent of ${\bf k}$,
special attention should be given to the transformation rules. In
particular, for the spin density vector we have \beq \vec{\cal
S}((k_x,k_y, k_z), ~ {\bf r}) = - \vec{\cal S}((-k_x,-k_y, k_z), ~
{\bf r}), \eeq i.e. the contributions to the matrix element
(\ref{matrixel}) from wave-vectors with opposite components parallel
to the surface have the same length and opposite orientations.
Consequently, in the ground-state the spin-density vanishes.
However, it is possible to generate a non-vanishing spin-density if
the number of particles moving in one direction is different from
the number of particles moving in the opposite direction, i.e. in
the presence of a charge current parallel to the boundary.

Let us assume now that the system is characterized by a preferential
direction of motion for the carriers imposed by an external current
flowing in the x-direction and having the average current density
$\langle j_x\rangle$. We reiterate that if the system is perfectly
clean and an external electric field is present, then  the particles
will just accelerate indefinitely and the spin density will be
time-dependent. This scenario is not considered here. Below we study
the situations when the equilibrium current appears either due to a
voltage drop which occurs entirely in the contacts, or because  some
disorder is present. In the latter case, we study only the
lengthscales
much smaller than the mean-free path (the physics at larger
length-scales in disordered systems will be discussed in
Sec.~\ref{Sec:Disorder}).
 We can associate the current with a drift
velocity ${\bf v}_0 = (\Delta k/m_H, 0, 0)$ with $\Delta k\ll k_F$,
and assume that the distribution function for the system changes
from $n^0({\bf k}) = \Theta(\epsilon_F - \epsilon_{\bf k})$ to
$n({\bf k}) = \Theta(\epsilon_F - \epsilon_{\bf k+\Delta k})$, where
$\Theta(x)$ is the step function, $\epsilon_{\bf k} = k^2/2m_H$ are
the energies of the eigenstates, and $\epsilon_F$ is the Fermi
energy corresponding to the Fermi wave-vector $k_F$ related to  the
density of particles, $n = k_F^3(1 + \xi^{\frac{3}{2}})/(6\pi^2)$.
Using the relation (\ref{sumint}) of transforming the sums over
${\bf k}$ into integrals, we have \beq \langle j_x \rangle =
\frac{1}{4\pi^3}\frac{e\hbar}{m_H}\left\{
\int_{\theta<\theta_c}d^3k~ \left[ 1+R_{LH}({\bf k}) \right] k_x
n({\bf k})+ \int_{\theta>\theta_c}d^3k~k_x n({\bf k}) \right\}
 = \frac{e\hbar}{m_H} \frac{1+\xi^{\frac{5}{2}}}{1+\xi^{\frac{3}{2}}}~n\Delta k,
\eeq where n is the average particle density of the system and we
have taken into account that the drift velocity is much smaller than
the Fermi velocity. In the presence of the external current, the
$y$-component of the spin density becomes non-zero and, using the
definition (\ref{jdens}) together with Eq.~(\ref{sumint}), we have
\beqa \langle S_y\rangle({\bf r}) &=&  \frac{1}{(2\pi)^3} \left\{
\sum_{\sigma} \int_{\theta<\theta_c}d^3k~\left[\mu_1(\theta){\cal
S}_y^{1\sigma}(\theta, {\bf r}) + \mu_2(\theta){\cal
S}_y^{2\sigma}(\theta, {\bf r})\right]
\cos(\phi) ~n({\bf k})\right. \label{expjy} \\
&+& \left.\int_{\theta>\theta_c}d^3k~\left[{\cal S}_y^{1}(\theta,
{\bf r}) + {\cal S}_y^{2}(\theta, {\bf r})\right]\cos(\phi) ~n_{\bf
k}\right\} = \frac{k_F^2\Delta
k}{8\pi^2}\left\{\int_0^{\theta_c}d\theta~{\cal S}_{<}(\theta, z) +
\int_{\theta_c}^{\pi/2}d\theta~{\cal S}_{>}(\theta, z)\right\},
\nonumber \eeqa where $\mu_i$ is the effective degeneracy of the
eigenstate, $n({\bf k}) = \Theta(\epsilon_F - \epsilon_{\bf k+\Delta
k})$, $\phi$ is the angle between the x-axis and the projection of
the wave-vector on the x-y plane, and the factor $\cos(\phi)$ is due
to the fact that ${\cal S}_y^{\alpha}(\theta, {\bf r})$ is
calculated in the local coordinate system. Finally, the functions
integrated over the angle $\theta$ in the last member of Eq.
(\ref{expjy}) are \beq {\cal S}_{<}(\theta, z) =
2\left[\mu_1(\theta){\cal S}_y^{1+}(\theta, z) + \mu_2(\theta){\cal
S}_y^{2+}(\theta, z)\right]\sin^2(\theta) ~~~~~~\mbox{and}~~~~~~
{\cal S}_{>}(\theta, z) = \left[{\cal S}_y^{1}(\theta, z) + {\cal
S}_y^{2}(\theta, z)\right]\sin^2(\theta). \eeq As expected, the spin
density is proportional to $\Delta k$, i.e. to the average density
of the external current. So far, we have reduced the expression for
the spin density $\langle S_y\rangle({\bf r})$ to a sum of one
dimensional integrals over the incident angle $\theta$. Furthermore,
we can write explicitly the z-dependence of ${\cal S}(\theta, z)$
using the expressions for the wave-functions and spinors as well as
the symmetry properties of the coefficients. We obtain \beqa
{\cal S}_{<}(\theta, z) &=& B(\theta)\left\{ \sin[2 \tilde{k}_z z] + \sin[2 \tilde{q}_z z] - 2\sin[(\tilde{k}_z+\tilde{q}_z)z] \right\}, \nonumber \\
{\cal S}_{>}(\theta, z) &=& B(\theta)\left\{ \sin[2 \tilde{k}_z z]
-2\sin[\tilde{k}_z z] e^{-\tilde{Q} z} \right\} +   C(\theta)\left\{
\cos[2 \tilde{k}_z z] + e^{-2 \tilde{Q} z} -2 \cos[\tilde{k}_z z]
e^{-\tilde{Q} z} \right\}, \label{Jthz} \eeqa where the
wave-vectors are proportional to $k_F$ and depend on the incident
angle $\theta$. Explicitly we have \beqa
\tilde{k}_z &=& k_F \cos(\theta), \nonumber \\
\tilde{q}_z &=& k_F \sqrt{\xi - \sin^2(\theta)},~~~~~~~~~~~~~~\theta<\theta_c,  \label{kqQ} \\
\tilde{Q}~  &=& k_F \sqrt{\sin^2(\theta) - \xi},~~~~~~~~~~~~~~\theta>\theta_c. \nonumber
\eeqa
The complete information about the spin density is contained in the
 coefficients $B(\theta)$ and $C(\theta)$. These quantities  can be
expressed in terms of the coefficients $c_i^{\alpha}$ of the
symmetric eigenfunctions (\ref{psisyp}) and (\ref{psisymm2}).
Explicitly, we have for $\theta<\theta_c$: \beq B(\theta) =
-6\left\{\mu_1\left[(c_1^1)^2 -(c_2^1)^2 \right] +
\mu_2\left[(c_1^2)^2 -(c_2^2)^2
\right]\right\}\cos(\theta)\sin^4(\theta) -12\left[\mu_1 c_1^1c_2^1
+ \mu_2 c_2^1c_2^2  \right]\cos^2(\theta)\sin^3(\theta),
\label{Bthet} \eeq where $\mu_i$ are the effective degeneracies
given by equation (\ref{mueff}) and the notations for the
coefficients $c_i^{\alpha}$ are taken from Table \ref{TBL2}.
Similarly, for $\theta>\theta_c$ we have \beqa
B(\theta) &=& -6\mbox{Re}\left[(c_1^1)^2 + (c_1^2)^2\right]\cos(\theta)\sin^4(\theta) + 6\mbox{Re}\left[i(c_1^1)^2 - i(c_1^2)^2\right]\cos^2(\theta)\sin^3(\theta)  \nonumber \\
C(\theta) &=& 6\mbox{Im}\left[(c_1^1)^2 +
(c_1^2)^2\right]\cos(\theta)\sin^4(\theta) -
6\mbox{Im}\left[i(c_1^1)^2 -
i(c_1^2)^2\right]\cos^2(\theta)\sin^3(\theta),   \label{BCthet}
\eeqa where the notations for the coefficients $c_i^{\alpha}$ are
taken from Table \ref{TBL4}. The explicit dependence of the
coefficients B and C on $\theta$ is shown in Fig.~\ref{FIG5} for
several values of the mass ratio $\xi$.
\begin{figure}
\begin{center}
\includegraphics[width=0.6\textwidth]{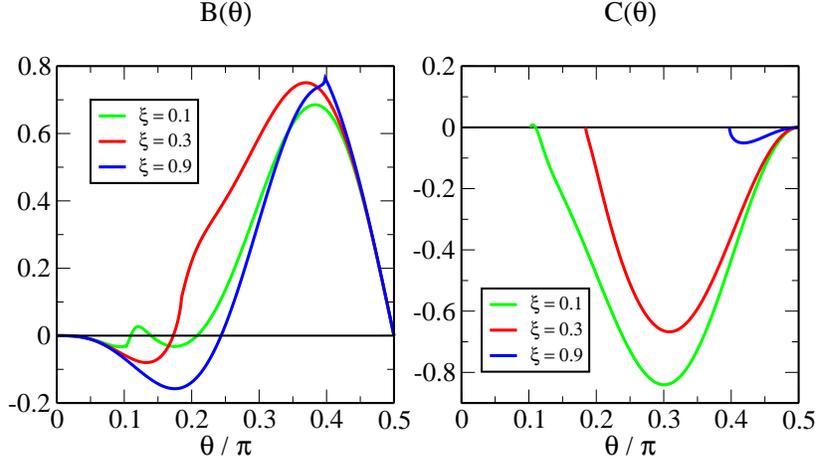}
\caption{(Color online) Angle dependence of the coefficients for the spin density defined by equations (\ref{Bthet}) and (\ref{BCthet}). The critical angles corresponding to the values of the mass ratio $\xi = 0.1, ~0.3, ~0.9$ are $\theta_c = 0.1024 \pi, ~0.1845 \pi$ and $0.3975 \pi$, respectively.}
\label{FIG5}
\end{center}
\end{figure}
Notice that the coefficient $C(\theta)$ vanishes at the critical
angle  $\theta = \theta_c$ and at $\pi/2$, while $B(\theta)$
vanishes at $\theta=0, ~\pi/2$ and has a singularity at the critical
angle characterized by a discontinuity of the first derivative.
Knowing these coefficients, we can perform the integral over the
angle $\theta$  in Eq.~(\ref{expjy}) and obtain the spin density.
The result for $\xi=0.3$ is shown in Fig.~\ref{FIG6} revealing
 several
interesting features. It confirms that the current-induced
spin density is indeed non-zero. In fact the spin density
 oscillates about zero with
an amplitude that vanishes as we move away from the boundary. The
oscillations are characterized by several  wave-lengths, as shown by
the beat phenomenon in the inset of Fig.~\ref{FIG6}. On the other
hand, the amplitude of the oscillations decreases as $1/z^2$ at
large distances. All these features are  contained
 in the properties of the coefficients $B(\theta)$
and $C(\theta)$, in particular in their behavior in the vicinity of
the singular points $\theta =0$, $\theta_c$ and $\pi/2$. Below, we
will
 analyze in detail the connection between these singular points and
the oscillating properties of the spin density.

\subsection{The asymptotic behavior of the spin density}
\label{Sec:Asymptotics}

The amplitude and the period of the spin density oscillations far
from the boundary are completely determined by the singular points
of the coefficients $B(\theta)$ and $C(\theta)$. To prove this, let
us consider as an example one of the terms from Eq.~(\ref{Jthz}) and
re-write the corresponding contribution to the spin density  as a
Fourier transform,
\beq \int_0^{\theta_c} d\theta~ B(\theta)\sin(2
\tilde{k}_z z) = \int_{-\infty}^{\infty} dp~ b(p) \sin(pz),
\label{Four}
\eeq
where
\beq b(p) = \left\{\begin{array}{ll}
\frac{B\left(\arccos\left(\frac{p}{2k_F}\right)\right)}{\sqrt{4-p^2}}~~~~~~~~~~~\mbox{if}~~~0\leq p \leq 2\sqrt{1-\xi}, \\
0 ~~~~~~~~~~~~~~~~~~~~~~~~~~~~~~~~~~~~\mbox{otherwise}.
\end{array}\right.  \label{discont}
\eeq If the Fourier coefficient $b(p)$ has a singularity at $p_0\neq
0$ characterized by a discontinuity $\Delta b^{(n)}(p_0)$ in the
derivative of order n, the large-z behavior of the integral
(\ref{Four}) is $\Delta b^{(n)}(p_0)~\cos(p_0 z+n \pi/2)/z^{n+1}$,
i.e. it oscillates with a period $\lambda = 2\pi/p_0$ and an
amplitude which is proportional to the discontinuity in the
derivative and
decays as $1/z^{n+1}$. Similar relations can be established for
cases when the discontinuity in the derivative is not finite. For
example,  when the singularity in proportional to $\sqrt{p-p_0}$,
the asymptotic behavior becomes  $\cos(p_0 z+ \pi/4)/z^{3/2}$.
Returning now to the analysis of the contributions to the spin
density coming from Eq.~(\ref{Jthz}), we notice that the possible
singularities are given by the angles $\theta \in \{\pi/2,~\theta_c,
~0\}$. As $\tilde{k}_z(\theta=\pi/2) = 0$, there is no large-z
contribution to the spin density coming from the states with the
incident angle $\theta=\pi/2$. In other words, the modes propagating
 parallel to the boundary do not contribute to the spin density
at positions far from the surface of the system. On the other hand,
the
\begin{figure}
\begin{center}
\includegraphics[width=0.5\textwidth]{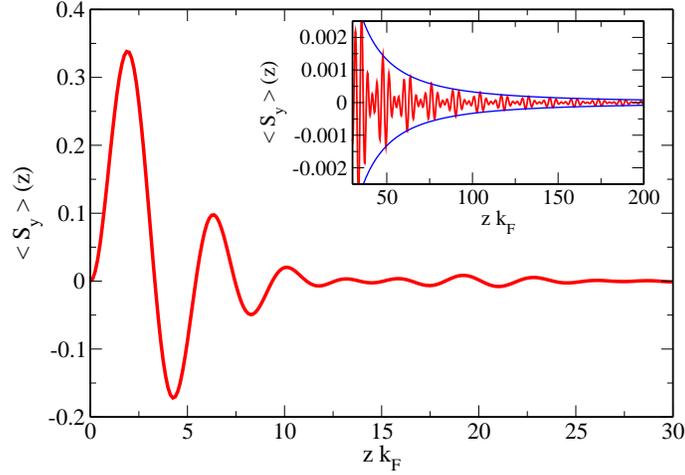}
\caption{(Color online) Position dependence of the current-induced
spin density $\langle S_y\rangle$ expressed in units of $k_F^2
\Delta k/8 \pi^2$ for a mass ratio $\xi = 0.3$ relevant for a GaAs
system. The inset shows the
large-z behavior characterized by a decay of the oscillations
proportional to $1/z^2$ (blue lines).} \label{FIG6}
\end{center}
\end{figure}
critical angle $\theta_c$ requires a detailed evaluation. We should
remind that, as  shown in the inset of Fig.~\ref{FIG6}, the spin
density oscillation decay as $1/z^2$. From Eq.~(\ref{Jthz}) we
observe that there are eight different terms  that contribute to the
spin density, three coming from  ${\cal S}_{<}$ and five coming from
${\cal S}_{>}$. However, one can  show that the contributions coming
from the last two terms proportional to $C(\theta)$ decay faster
that $1/z^2$ and, consequently, can be neglected. The remaining six
terms will give contributions that decay as $1/z^2$ or slower. For
example, the first term $B(\theta) \sin(\tilde{k}_z z)$ with
$\theta<\theta_c$ generates a contribution proportional to
$B(\theta_c) \cos(2\sqrt{1-\xi}z)/z$ due to the fact that
$B(\theta_c)\neq 0$. However, this contribution is exactly canceled
by the corresponding contribution coming from the fourth term,
$B(\theta) \sin(\tilde{k}_z z)$ with $\theta>\theta_c$. Similarly,
the $\sqrt{\theta_c-\theta}$ term from  $B(\theta)$ generate the
contribution $\cos(2\sqrt{1-\xi}z +\pi/4)/z^{3/2}$, which is exactly
canceled by the  contribution generated by $C(\theta)$. In fact, one
can show that there is no net contribution   to the asymptotic spin
density of order up to ${\cal O}(1/z^2)$ coming from the
critical angle. This exact cancellation is embedded in the properties
of the coefficients $B(\theta)$ and $C(\theta)$ in the vicinity of
$\theta_c$. Explicitly, for $\theta \approx \theta_c$ we have \beqa
B(\theta) &=& \beta_0 + \beta_{1/2}\sqrt{\theta_c-\theta} + \beta_1 (\theta_c-\theta) + {\cal O}\left((\theta_c-\theta)^{3/2} \right), ~~~~~~~~~\theta<\theta_c  \nonumber \\
B(\theta) &=& \beta_0  -  \beta_1 (\theta-\theta_c) + {\cal O}\left((\theta-\theta_c)^{2} \right), ~~~~~~~~~~~~~\theta>\theta_c  \\
C(\theta) &=& \beta_{1/2}\sqrt{\theta-\theta_c}  + {\cal
O}\left((\theta-\theta_c)^{3/2} \right),
~~~~~~~~~~~~~\theta>\theta_c.  \nonumber \eeqa We conclude that the
asymptotic behavior of the spin density is determined entirely by
the modes propagating along the direction perpendicular to the
boundary. The characteristic wave-vectors corresponding to the first
three terms in Eq.~(\ref{Jthz}) are $2\tilde{k}_z(0) = 2k_F$,
$2\tilde{q}_z(0) = 2k_F\sqrt{\xi}$ and $\tilde{k}_z(0) +
\tilde{q}_z(0) = k_F(1+\sqrt{\xi})$. On the other hand, the
amplitude
 is determined by the small angle behavior of the coefficient B,
namely
\beq B(\theta) = -3 \theta^3 + {\cal O}(\theta^5),
~~~~~~~~~\mbox{for}~~\theta \rightarrow 0. \eeq Writing each term as
a Fourier transform and using the properties discussed above in
connection with equations (\ref{Four}-\ref{discont}) we obtain the
following expression for the spin density far from the boundary \beq
\langle S_y \rangle (z) = \frac{\Delta k}{8
\pi^2}~\frac{3}{2z^2}~\left\{\sin\left(2k_F z\right) +
\xi\sin\left(2k_F \sqrt{\xi} z \right)
-\frac{8\sqrt{\xi}}{(1+\sqrt{\xi})^2}\sin\left[k_F(1+ \sqrt{\xi})
z\right]\right\}, ~~~~~~~~~~~~~~zk_F\gg 1, \label{asymp}
\eeq
where the Fermi wave-vector is determined by the density and the
spin-orbit coupling, $k_{\rm F} = [6\pi^2 n/(1+\xi^{3/2})]^{1/3}$.
We note that the result (\ref{asymp}) is asymptotically exact.

\begin{figure}
\begin{center}
\includegraphics[width=0.5\textwidth]{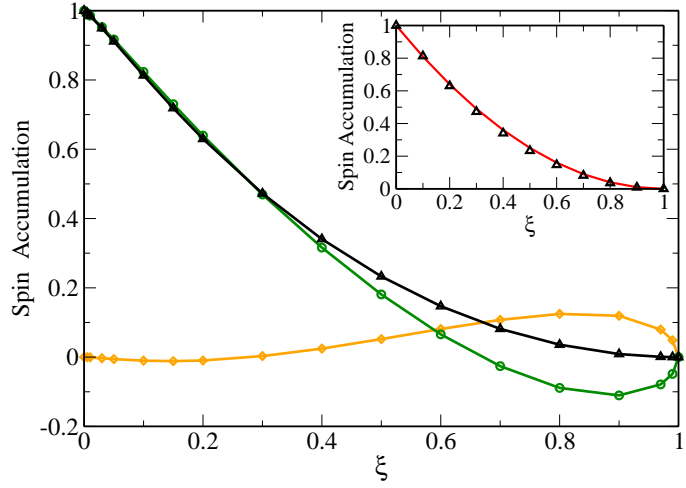}
\caption{(Color online) Total spin accumulation in units of $k_F
\Delta k/8 \pi^2$ as a function of the mass ratio $\xi$ (black line
with triangles). The contributions from angles $\theta < \theta_c$
(orange line with diamonds) and $\theta > \theta_c$ (green  line
with circles) are shown separately. The inset represents a fit of
the total spin accumulation (triangles) with the function
$(1-\xi)^2$ (continuous red line).} \label{FIG7}
\end{center}
\end{figure}

\subsection{The spin accumulation}
\label{Sec:Accumulation}

An important question is whether there is a net spin accumulation in
the system. To answer this question, we first define the spin
accumulation as an integral of the spin density with respect to the
distance from the boundary \beq \overline{S}_y(z) = \int_0^z \langle
S_y \rangle (\zeta) ~d\zeta. \eeq Here $\overline{S}_y(z)$ is the
spin accumulation at a distance z from the boundary, more precisely
the spin density per unit area. The total spin accumulation is
$\overline{S}_y = \overline{S}_y(\infty)$. To determine the total
spin accumulation it is convenient to perform first the integration
over the coordinate z for each of the terms contributing to ${\cal
S}(\theta, z)$.  We obtain for the total spin accumulation the
expression \beq \overline{S}_y =
\int_0^{\theta_c}d\theta~B(\theta)\left[\frac{1}{2 \tilde{k}_z} +
\frac{1}{2 \tilde{q}_z} - \frac{2}{\tilde{k}_z+\tilde{q}_z} \right]
+ \int_{\theta_c}^{\pi/2} d\theta~\left\{B(\theta)\left[\frac{1}{2
\tilde{k}_z} + \frac{\tilde{k}_z}{\tilde{k}_z^2 +\tilde{Q}^2}\right]
+ C(\theta)\left[\frac{1}{2\tilde{Q}} -
\frac{2\tilde{Q}}{\tilde{k}_z^2 +\tilde{Q}^2}\right] \right\}, \eeq
where the dependence on $\theta$ of the wave-vectors $\tilde{k}_z$,
 $\tilde{q}_z$ and  $\tilde{Q}$ is given by Eq.~(\ref{kqQ}). Finally,
 the integration over the angle is performed numerically. The
result for the spin accumulation as a function of the mass ratio
$\xi = m_L/m_H$ is shown in Fig.~\ref{FIG7}. The spin accumulation
decreases monotonically with $\xi$ and vanishes  in the absence of
spin-orbit coupling ($\xi = 1$). We have also calculated separately
the contributions from the angles smaller than the critical angle
$\theta_c$ (orange line in Fig.~\ref{FIG7}) and the angles larger
than $\theta_c$ (green line). For strong to intermediate values of
the spin-orbit interaction  practically the entire contribution to
the spin accumulation comes from states with $\theta
> \theta_c$, i.e. the states containing localized modes. In contrast, in
the weak-coupling regime ($\xi \sim 1$), the two contributions are
comparable and have opposite signs. Finally, we notice that  the
$\xi$-dependence of the total spin accumulation can be perfectly
fitted  as (see the inset in Fig.~\ref{FIG7}) \beq \label{guess}
\overline{S}_y \approx \frac{k_F\Delta
k}{8\pi^2}\left(1-\xi\right)^2. \eeq

The spin accumulation as a function of distance has also been
calculated numerically and the result is shown in Fig.~\ref{FIG8}.
\begin{figure}
\begin{center}
\includegraphics[width=0.5\textwidth]{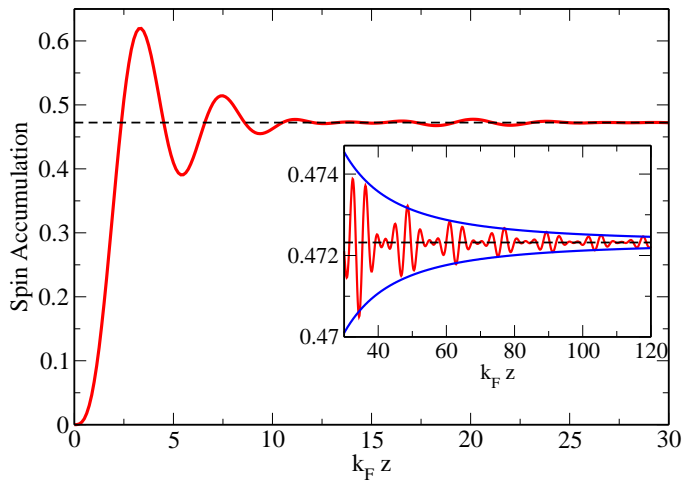}
\caption{(Color online) Current-induced spin accumulation in units
of $k_F \Delta k/8 \pi^2$ as a function of the distance from the
boundary. The curve was determined for $\xi=0.3$ by integrating the
data from Fig.~\ref{FIG6} with respect to z from zero to a certain
distance from the boundary. Notice that the  spins are localized
within a few Fermi lengths $\lambda_F = 2\pi/k_F$ off the boundary.
The dashed line represents the total spin accumulation
$\overline{S}_y$ using the same units. The inset shows the $1/z^2$
decay of the fluctuations about $\overline{S}_y$ at distances far
from the boundary.} \label{FIG8}
\end{center}
\end{figure}
We see that the spin accumulation is localized in the vicinity of
the boundary. The shortest distance from the boundary at which the
spin accumulation becomes equal to the total spin accumulation
 $\overline{S}_y$ is of  the order of $2/k_F$, while at larger
distances it oscillates about $\overline{S}_y$  with an amplitude
that decreases as we move away from the boundary. As shown in the
inset of Fig.~\ref{FIG8}, at large distances the oscillations of the
spin accumulation decay as $1/z^2$.

\section{The effects of disorder} \label{Sec:Disorder}

\subsection{The Friedel oscillations in a disordered system}
\label{Sec:FrDis} In this section we qualitatively discuss the
effects of disorder on the current-induced Friedel oscillations of
the spin density. To understand these effects it is useful to recall
the corresponding physics in the usual disordered Fermi systems:
Consider a perturbation  (such as a boundary or an impurity)
introduced in a Fermi gas. This perturbation leads to the Friedel
oscillations in the electron density, which in a clean case behaves
as $n(r) = A r^{-p(d)} \cos{\left( 2 p_{\rm F} r\right)}$, where $r$
is the distance from the perturbation, $A$ is a constant amplitude,
and the power $p(d) = (d+1)/2$, if the perturbation is a boundary
and $p(d) = d$, if the perturbation is point-like (i.e., a magnetic
impurity). What happens with the Friedel oscillations if we
introduce disorder? A natural quantity to calculate is the
perturbation-induced density averaged over disorder realizations. As
was shown by de Gennes,~\cite{deGennes} it decays exponentially as
$\left\langle n({\bf r}) \right\rangle \propto e^{-r/l}$, were $l$
is the mean free path. However, Zyuzin and
Spivak~\cite{Zuzin_spivak} later argued that this result is
misleading. Indeed, if we measure  the density at different points
but at the same distance away from the perturbation, we will get
different results (in magnitude and sign), which will strongly
depend on the local disorder realization (see also,
Refs.~[\onlinecite{Lerner},~\onlinecite{GVG}]). Therefore a more
meaningful quantity is the full probability distribution function of
density at a distance $r$ from the perturbation $P[n,r]$. One should
note that this probability distribution function is not known
explicitly even in the simplest cases and only its lowest moments
are available. The important result of Zyuzin and
Spivak~\cite{Zuzin_spivak}  (obtained in the context of RKKY
oscillations between magnetic impurities in disordered metals)
concerns the second moment, which was shown to decay as a power law,
$\left\langle n({\bf r}) n({\bf r}) \right\rangle \propto r^{-p(d)}$
just like in the clean case. The qualitative explanation of this
result is that disorder effectively introduces random phase shifts
$\delta(l)$ in the argument of the cosine in the Friedel
oscillations $\propto \cos{\left[2 p_{\rm F} r + \delta(l)\right]}$.
Averaging over disorder realizations leads effectively to the
vanishing density due to the oscillations in the cosine. However,
averaging of the cosine squared yields a finite result and restores
the power law behavior of the {\em typical} Friedel oscillation.

To connect the above-mentioned arguments with our case, we mention
that the higher moments of the probability distribution function of
the Friedel oscillations in the density are described
diagrammatically by bubble diagrams with all possible disorder
averagings between them. A careful treatment of these diagrams
reveals that the power-law (non-exponential) decay of the usual
Friedel oscillations is due to the gapless nature of the diffusons
of which these diagrams are built. Now let us return to the Friedel
oscillations in the spin density and spin accumulation, which were
discussed above in the clean case. A straightforward calculation
shows that the spin density averaged over disorder  decays
exponentially as \beq \label{<S>} \overline{{\bf S}({\bf r})}
\propto \int d\varepsilon\, {\rm Tr}\, \left[ \hat {\bf S} \hat
{\overline{G}} (\varepsilon; {\bf r},{\bf r}) \right] \propto
e^{-r/l}, \eeq where $l$ is the mean free path, $\hat G(\varepsilon;
{\bf r},{\bf r})$ is the Green's function which takes into account
the boundary, and overline implies disorder averaging. Higher
moments of the spin density are described by expressions, which
involve different components of the diffuson. In particular, the
second moment (correlator of spin densities) reads
\beq \label{<SS>}
\overline{S_\alpha ({\bf r}) S_\alpha ({\bf r})} \propto \int
d\varepsilon_1  \int d\varepsilon_2 \overline{ {\rm Tr}\, \left[
\hat S_\alpha \hat {G} (\varepsilon_1; {\bf r},{\bf r})\right] {\rm
Tr}\, \left[ \hat S_\alpha \hat {G} (\varepsilon_1; {\bf r},{\bf r})
\right]} \propto \frac{1}{r^4},
\eeq
We emphasize that due to the two traces present in
Eq.~(\ref{<SS>}), the correlation function includes contributions
from different components of the diffuson matrix. Most importantly
it contains a singlet component, which remains gapless even in a
spin-orbit coupled system (see below subsection~\ref{Sec:SpinRel}) .
Therefore, the {\em typical} spin density measured at a distance $r
\gg l$ from the boundary is expected to decay as a power law even if
disorder is present. This spin density is however random in sign and
can be estimated as $S_y ({\bf r}) \propto z^{-2} \sin\left[
\phi_r({\bf r}) \right]$, where $z$ is the distance from the
boundary and $\phi_r$ is a random phase. Due to this randomness, the
spin accumulation ({\em i.e.} the spin density averaged over the
length-scales much larger than the Fermi-wavelength) is expected to
decay exponentially as $\propto e^{-z/L_s}$, where $L_s$ is the spin
relaxation or spin diffusion
length.~\cite{SZhang,Anton,Mishchenko_etal} We note that in the
framework of the Luttinger model, the spin-orbit coupling is always
strong and therefore the spin relaxation length is expected to be
very short and of the order of the mean free path. We study the
related spin relaxation time in the following subsection.

\subsection{Spin relaxation time in the Luttinger model}
\label{Sec:SpinRel} For the sake of completeness, we calculate
explicitly the spin relaxation time and find  that it is very short
(of order $\tau$). Therefore, the spin relaxation length is short too
and the hydrodynamic diffusion approximation is not applicable for
spin transport in the Luttinger model. However, we believe that it
is likely that the result for the spin relaxation time itself is
quantitatively correct and describes time relaxation of a uniform
spin distribution.

To find the spin relaxation time, we construct the diffuson, which
can be obtained from a convolution of disorder-averaged Green's
functions [see Eqs.~(\ref{DIP}) and (\ref{Isigsig}) below]. For the
purpose of this section, we can ignore the boundary effects and use
the bulk Green's function in the calculations.
We consider the system in the presence of randomly distributed
impurities with short-range potential $V_{\rm imp}({\bf r}) = u_0
\sum_i \delta({\bf r} - {\bf r})$, and calculate the self-energy in
the Born approximation\cite{murak}.
We obtain for the retarded (advanced) self-energy the expression
\beq \Sigma_{{\bf k}\lambda, {\bf
k}^{\prime}\lambda^{\prime}}(\omega\pm 0^+) \approx \frac{\mp i\pi
n_iv_0^2\rho_F}{2}\delta_{{\bf k}{\bf k}^{\prime}}\delta_{\lambda
\lambda^{\prime}} = \frac{\mp i}{2\tau}\delta_{{\bf k}{\bf
k}^{\prime}}\delta_{\lambda \lambda^{\prime}}, \label{sigtau} \eeq
where $\rho_F = m_Hk^{(H)}_F(1+\xi^{3/2})/\pi^2$ is the density of
states at the Fermi energy, $n_{i}$ represents
the density of impurities,  and $\tau = 1/(\pi n_iv_0^2\rho_F)$ is
the mean scattering time. Notice that the self-energy in the first
Born approximation is momentum independent and diagonal in both
momentum and helicity. We consider here the quasi-classical limit,
when the mean-free paths for heavy- and light-holes are  much longer
than the corresponding Fermi lengths,
$l_{\lambda}=k^{(\lambda)}_F\tau/m_{\lambda}\gg
2\pi/k^{(\lambda)}_F$. Using  the expression (\ref{sigtau}) for the
self-energy, we obtain for the disorder-averaged retarded Green's
function the usual expression\cite{zhang} \beq {\hat
{\overline{G}}}^R_{\bf k}(\omega) = \left[f^R(k, \omega) +
\frac{5}{4}g^R(k, \omega) \right] \hat I - g^R(k, \omega)({\bf n_k}
\hat {\bf S})^2 = f^R(k, \omega) \hat I + g^R(k, \omega) ~{\bf
d}({\bf n_k})\hat {\bf \Gamma}, \label{Gbar} \eeq where ${\bf n_k} =
{\bf k}/|{\bf k}|$ is the unit vector directed along the momentum,
$\hat I$ represents the $4\times4$ unit matrix, the $\hat {\bf
S}$-matrices are given in appendix \ref{AppA}, and ${\bf d}({\bf
n_k})\hat {\bf \Gamma}=d_i({\bf n_k})\hat \Gamma_i$ represents the
product of a five-component unit vector that depends on  the
direction of ${\bf k}$. In what follows we use the technique
introduced by Zhang {\em et al.}, which involves the five generators
$\hat \Gamma_i$ of the SO(5) Clifford algebra\cite{zhang}.
Finally, the coefficients $f^R$ and $g^R$ are independent of the
direction of the momentum and read \beq \left\{\begin{array}{cc}
f^R(k, \omega) \\
g^R(k, \omega)
\end{array}\right\} =\frac{1}{2}\left[\frac{1}{\omega+\mu-\frac{k^2}{2m_H} + \frac{i}{2\tau}} \pm \frac{1}{\omega+\mu-\frac{k^2}{2m_L} + \frac{i}{2\tau}}\right],
\eeq
where $\mu$ is the chemical potential. Similar expressions can
be obtained for the advanced Green's function with
$f^A(k, \omega) =[f^R(k, \omega)]^*$ and
$g^A(k, \omega) =[g^R(k, \omega)]^*$.

We are ready now to  calculate the kernel of the diffusion equation,
\beq \breve{\cal D} = [\hat I\otimes \hat I- \breve \Pi]^{-1}
\label{DIP} \eeq where $\breve {\cal D}$  and $\breve \Pi$ are
$16\times 16$ matrices labeled by two pairs of ``spin'' indices
$\sigma \in\{3/2,~1/2, ~-1/2, ~-3/2\}$, and $\hat I$ is the
$4\times4$ unit
matrix. The polarizability can be expressed as a convolution of two
Green's functions, \beq \Pi_{\sigma_1\sigma_2,
\sigma_3\sigma_4}(\omega, {\bf q}) =
\frac{2}{\pi\rho_F\tau}\int\frac{d^3k}{(2\pi)^3}~\left[\overline{G}_{{\bf
k}+{\bf q}/2}^R(\omega/2)\right]_{\sigma_3\sigma_1}
\left[\overline{G}_{{\bf k}-{\bf
q}/2}^A(-\omega/2)\right]_{\sigma_2\sigma_4}. \label{Isigsig} \eeq
Considering now the limit $\omega \ll \epsilon_F$ and ${\bf q}=0$
 in Eq.~(\ref{Isigsig}) and performing the integral over momenta
we obtain \beq \breve \Pi(\omega, 0) =
\frac{1}{2}(1+i\omega\tau)\left[\hat I\otimes \hat I +
\frac{1}{5}\sum_i\hat  \Gamma_i \otimes \hat  \Gamma_i \right] +
{\cal F}(\xi)(1+i\omega\tau)\left[\hat  I\otimes \hat I -
\frac{1}{5}\sum_i \hat  \Gamma_i\otimes \hat \Gamma_i \right],
\label{Pi0} \eeq where $\hat I$ is the $4\times4$ unit matrix, $\hat
\Gamma_i$ are the gamma-matrices defined in [\onlinecite{zhang}],
and the coefficient ${\cal F}$ depends on the strength of the
spin-orbit coupling. Explicitly we have \beq {\cal F}(\xi) =
\frac{1}{4(\epsilon_F\tau)^2}\frac{\xi}{1+\xi^3/2}\frac{(1+\sqrt{\xi})
(1+\xi)-\frac{1}{2}(1-\sqrt{\xi})(1-\xi)}{(1-\xi)^2+\frac{1}{4(\epsilon_F\tau)^2}(1+\xi)^2},
\eeq where $\xi = m_L/m_H = (1-2\gamma)/(1+2\gamma)$ is the
light-hole to heavy-hole mass ratio and  $\epsilon_F$ is the Fermi
energy. Notice
 that $\epsilon_F\tau \gg 1$ and, consequently, the coefficient
${\cal F}$ is small, ${\cal F} \sim {\cal O}(1/(\epsilon_F\tau)^2)$,
for any value of the spin-orbit coupling satisfying the inequality
 $\gamma \gg 1/(4\epsilon_F\tau)$, which is always the case for
physically relevant parameters. We recall that the small spin-orbit
coupling limit is not physically consistent within the Luttinger
model (if $\gamma \to 0$, additional bands have to be
considered), although mathematically well-defined. Formally, we have
${\cal F}(1) = 1/2$ and we recover the standard result for systems
without spin-orbit coupling. If we substitute now the expression
(\ref{Pi0}) of the polarizability into Eq.~(\ref{DIP}), we obtain
for the diffuson \beq \breve {\cal D}^{-1} = \left[\frac{1}{2}-{\cal
F}(\xi)\right]\left\{\hat I\otimes \hat I - \frac{1}{5}\sum_i\hat
\Gamma_i\otimes \hat \Gamma_i \right\}
-i\omega\tau\left\{\left[\frac{1}{2}+ {\cal F}(\xi) \right] \hat
I\otimes \hat I +  \left[\frac{1}{2}- {\cal F}(\xi)
\right]\frac{1}{5}\sum_i\hat \Gamma_i\otimes \hat \Gamma_i \right\}
\label{calD} \eeq The final step in our evaluation is to write the
diffuson in a different representation that has a more transparent
physical meaning.  Explicitly, we define \beq \tilde{\cal D}_{ab} =
 M_{\sigma_1\sigma_2}^a{\cal D}_{\sigma_1\sigma_2,
\sigma_3\sigma_4}M_{\sigma_3\sigma_4}^b,        \label{mtpl} \eeq
where $M_{\sigma_1\sigma_2}^a$ are  $4\times 4$ matrices and
summation over repeating indices is implied. The M-matrices satisfy
the resolution of identity
$M_{\sigma_1\sigma_2}^aM_{\sigma_2^{\prime}\sigma_1^{\prime}}^a =
\delta_{\sigma_1 \sigma_1^{\prime}} \delta_{\sigma_2
\sigma_2^{\prime}}$ and have the property  Tr$[(\hat M^a)^2] =1$. In
particular, $M^0_{\sigma_1\sigma_2}=\delta_{\sigma_1\sigma_2}/2$
represents the charge channel, while $\hat M^1=\hat S_x/\sqrt{5}$,
$\hat M^2=\hat S_y\sqrt{5}$, and  $\hat M^3=\hat S_z\sqrt{5}$
correspond to the spin channels. Notice that the original quantity
is a $16\times 16$ matrix and, consequently, there will be 12 other
channels, in addition to charge and spin, corresponding to higher
magnetic moments.~\cite{wink} In fact Eq.~(\ref{mtpl}) can be viewed
as a multipole expansion of the density matrix\cite{wink} with the
additional channels associated with the five $\Gamma$-matrices which
are quadratic combination of the spin matrices\cite{zhang}
(``quadrupole moments'') and
 seven linearly independent matrices containing cubic
terms\cite{zhang} (``octupole moments''). Introducing the result
(\ref{calD}) in Eq.~(\ref{mtpl}) we obtain the kernel of the
diffusion equation in the new representation \beq \tilde{\cal
D}^{-1}_{ab}(\omega, 0) = \left[\frac{1}{2}-{\cal F}(\xi) \right]
\alpha_a \delta _{ab} -i\omega\tau \left[ \frac{1}{2}\beta_a + {\cal
F}(\xi)\alpha_a\right]\delta _{ab},    \label{diffus} \eeq with \beq
\alpha_a = 1-\frac{1}{5}\sum_i\mbox{Tr}[\hat M^a\hat \Gamma_i \hat
M^a\hat \Gamma_i], ~~~~~~~~~~~~~~~~\beta_a =
1+\frac{1}{5}\sum_i\mbox{Tr}[\hat M^a\hat \Gamma_i \hat M^a\hat
\Gamma_i]. \eeq Finally, from equation (\ref{diffus}) we evaluate
the relaxation times \beq \frac{1}{\tau_a} =
\frac{\left[\frac{1}{2}-{\cal F}(\xi) \right]
\alpha_a}{\frac{1}{2}\beta_a + {\cal
F}(\xi)\alpha_a}~\frac{1}{\tau}\approx
\frac{\alpha_a}{\beta_a}~\frac{1}{\tau}, \eeq where for the
approximation we considered physically relevant parameters
satisfying $1-\xi \gg 1/(\epsilon_F\tau)$. Formally, in the limit of
vanishing spin-orbit interaction $\xi=1$ we have ${\cal F}(1)=1/2$
and all the relaxation times become infinite. However, in the
physically relevant regime, relaxation is naturally absent only in
the charge channel as $\alpha_0 = 0$ and $\beta_0 = 2$. The spin
relaxation times are
$$
\tau_x=\tau_y=\tau_z=\frac{3}{2}\tau.
$$
The relaxation times for the ``quadrupole'' and ``octupole''
channels are
$$
\tau_q = \frac{\tau}{4}\,\,\,\, \mbox{and}\,\,\,\, \tau_o =
\frac{3}{2} \tau.
$$
Notice that all these relaxation times do not depend on the
spin-orbit coupling [if $1- m_L/m_H \gg \left(\epsilon_F
\tau\right)^{-1}$] and are comparable with the scattering time.
Thus, these channels can not be described within the diffusion
approximation.

\section{Conclusion}

We have found that the boundary spin Hall effect in a clean
hole-doped semiconductor  can be viewed as current-induced Friedel
oscillations of the spin density with  a net  spin accumulation
originating, in particular, from the localized surface states of the
light-holes. Although, the explicit results derived in this paper
are specific to the Luttinger model, the main qualitative
conclusions generalize to other spin-orbit coupled
systems.~\cite{Rashba_SO,Dresselhaus_SO} Indeed, the existence of a
Fermi surface in a metal or a semiconductor inevitably leads to the
Friedel oscillations if a perturbation (such as a boundary or an
impurity) is introduced in the Fermi gas. In the presence of a
spin-orbit coupling, there are multiple Fermi surfaces and therefore
there should be Friedel oscillations with multiple periods. It is
also clear that the existence of multiple Fermi surfaces should lead
to the appearance of surface states: As one can see from the
derivation of Sec.~\ref{Sec:Eignestates}, these states occur due to
the impossibility to satisfy the energy and momentum conservation
laws simultaneously if a majority carrier is reflected from the
boundary having a large angle of incidence. This results in the
appearance of localized surface states of minority
carriers.~\cite{Usaj,Shekhter} We note that since different types of
carriers appear due to the spin-orbit splitting and have different
spin properties, the surface states of minority carriers will always
lead to a boundary spin accumulation if there is an equilibrium
charge current present. These boundary states may also be important
in the context of spin diffusion in disordered systems with weak
spin-orbit coupling. Indeed, it has been shown that the boundary
spin accumulation in such systems is extremely sensitive to the
boundary
conditions.~\cite{New_Awschalom,TFZDS,GBSDS,Malsh2,Intr_edges} The
existence of the surface states may affect boundary conditions,
since the ``localization'' of the minority carriers at the boundary
would mimic a partially transparent interface. We note here that
this effect is definitely non-perturbative and cannot be captured
within expansions in powers of  spin-orbit coupling, which is
usually used in the context of spin diffusion.

It seems plausible that the existence of the Tamm-like surface states
can be connected to the Fermi surface Berry's phase structure in
spin-orbit coupled systems, but this connection remains unclear at
this stage. It is also unclear whether the spin Hall conductivity
can be related to any of the observable properties in the spin Hall
type experiment. In our analysis, we have not been able to identify
such quantities, although the orientation of the spin accumulated
near the boundary  qualitatively agrees with the predictions of the
spin-current theory.
The main inconsistency is that the spin Hall
conductivity tends to a finite constant if the spin-orbit coupling
is small, while the spin accumulation calculated here depends
strongly on the spin-orbit coupling $\gamma$ and vanishes if the
latter goes to zero.

We emphasize that our qualitative analysis is applicable only at
length-scales much smaller than the mean-free path. At larger length
scales different mechanisms of spin accumulation come into play. One
such mechanism is due to the Dyakonov-Perel-like spin
relaxation,~\cite{DP1,DP2,Edelstein} which together with the
spin-charge coupling may provide another contribution to spin
accumulation. However, as shown in Sec.~\ref{Sec:Disorder}, the spin
relaxation times are always too short in the framework of the
Luttinger model and therefore the above-mentioned disorder-induced
spin accumulation will decay exponentially at  distances of order
mean free-path. In particular, this implies that the semiclassical
diffusion approach is not applicable to the problem.

In summary, we have constructed an  exact orthonormal basis for the
isotropic Luttinger model in half-space. It contains the usual
incident and reflected waves as well as novel localized light-hole
states. These surface states contribute to the
 net spin accumulation in response to a charge current.
To experimentally observe the
predicted  surface states, it may be useful to
consider a set-up in which a three-dimensional semiconductor (e.g.,
GaAs) exists in a close proximity to a low-dimensional system
(which does not necessarily have a strong spin-orbit coupling) with
a contact near the three-dimensional boundary. In this case,
applying current to the system would result in spin injection into
the low-dimensional system, which should be directly observable by
the means of the Kerr effect. The experimental verification of the
predicted spin-polarized surface states in hole-doped semiconductors
is clearly called for.

\begin{acknowledgments}
The authors are grateful to Anton Burkov, Alexei Kaminski, Boris Spivak, and Maxim
Vavilov for valuable insights.
\end{acknowledgments}

\appendix

\section{Spin 3/2 matrices and spinors} \label{AppA}

The expressions for the S matrices in a basis with the
total angular momentum parallel to the z axis are:
\beq
\hat S_x = \left(
\begin{array}{cccc}
~&\frac{\sqrt{3}}{2}&~&~\\ \frac{\sqrt{3}}{2}&~&1&~\\
~&1&~&\frac{\sqrt{3}}{2}\\~&~&\frac{\sqrt{3}}{2}&~
\end{array}
\right),~~~~~
\hat S_y = \left(
\begin{array}{cccc}
~&-\frac{\sqrt{3}}{2}i&~&~\\ \frac{\sqrt{3}}{2}i&~&-i&~\\
~&i&~&-\frac{\sqrt{3}}{2}i\\~&~&\frac{\sqrt{3}}{2}i&~
\end{array}
\right),~~~~~
\hat S_z = \left(
\begin{array}{cccc}
\frac32&~&~&~\\~&\frac12&~&~\\~&~&-\frac12&~\\~&~&~&-\frac32
\end{array}
\right).
\eeq  \label{A1}

For a translation invariant system the eigenvectors of the Luttinger
Hamiltonian are given by equation (\ref{spinor}). The four component
spinors $U_{\lambda}({\bf n_k})$ depend on the orientation, not the
magnitude, of the wave vector, so each of them can be uniquely
described by two angles. For the purpose of this article it is
enough to know their expressions in a particular coordinate system
in which the $y$-component of the wave vector is zero, $k_y = 0$.
Consequently, the spinors will depend on one angle $\theta$, which
we choose as the angle between the z-axis and the  vector $-{\bf
k}$. Explicitly we have for $\theta < \theta_c$ \beqa U_{H+}(\theta)
&=& \left(
\begin{array}{cccc}
\sin^3(\frac{\theta}{2}) \\
\sqrt{3}\cos(\frac{\theta}{2})\sin^2(\frac{\theta}{2}) \\
\sqrt{3}\cos^2(\frac{\theta}{2})\sin(\frac{\theta}{2}) \\
\cos^3(\frac{\theta}{2})
\end{array}
\right),~~~~~~~~
U_{L+}(\theta) = \left(
\begin{array}{cccc}
-\sqrt{3}\cos(\frac{\theta}{2})\sin^2(\frac{\theta}{2}) \\
-1/2\left[1+3\cos(\theta)\right]\sin(\frac{\theta}{2}) \\
~~1/2\left[ 1-3\cos(\theta) \right]\cos(\frac{\theta}{2}) \\
~~\sqrt{3}\cos^2(\frac{\theta}{2})\sin(\frac{\theta}{2})
\end{array}
\right), \nonumber \\ \nonumber \\
U_{H-}(\theta) &=& ~U_{H+}(\pi+\theta), ~~~~~~~~~~~~~~~~~~~~~~
U_{L-}(\theta) = ~U_{L+}(\pi+\phi). \eeqa For an incident angle
$\theta > \theta_c$ there are no light-holes propagating in the z
direction, but localized solutions described by the spinors \beqa
V_1(\chi) &=&
\frac{1}{2\sqrt{2\left[1+\sin(\chi)+\sin^2(\chi)\right]}}
\left(\begin{array}{cccc} ~i\sqrt{3}\\ ~~~1 + 2\sin(\chi)\\
-i\left[1 + 2\sin(\chi)\right] \\ -\sqrt{3}
\end{array}\right), \nonumber \\
\nonumber \\
V_2(\chi) &=&
\frac{1}{2\sqrt{2\left[1-\sin(\chi)+\sin^2(\chi)\right]}}
\left(\begin{array}{cccc} -\sqrt{3}\\ -i\left[1 - 2\sin(\chi)\right]\\
~~~1 - 2\sin(\chi) \\ ~i\sqrt{3}
\end{array}\right).        \label{Vqchi}
\eeqa

\section{Scattering coefficients for an incident heavy-hole with helicity +3/2} \label{AppB}

The scattering coefficients $A_i$ and $B_i$ in Eq.
(\ref{psihp}) are functions of the incident angle $\theta$ and
can be obtained by imposing the boundary condition
$\psi_{\bf k}^{(H+)}(z=0) = 0$. Explicitly we have
\beq
\left(\begin{array}{ccc}
A_1(\theta) \\ A_2(\theta) \\ B_1(\theta) \\ B_2(\theta)
\end{array}\right) =
\frac{1}{5 + 3\cos[2(\phi(\theta)-\theta)]}
\left(\begin{array}{ccc}
-(4 - 3\cos[2\phi(\theta)] - \cos[2\theta])\sin[\theta] \\
~~~(4 + 3\cos[2\phi(\theta)] + \cos[2\theta])\cos[\theta] \\
\sqrt{3}(\sin[\frac{1}{2}(\phi(\theta) - 3\theta)] - 3\sin[\frac{1}{2}(3\phi(\theta) - \theta)]) \sin[2\theta] \\
\sqrt{3}(\cos[\frac{1}{2}(\phi(\theta) - 3\theta)] + 3\cos[\frac{1}{2}(3\phi(\theta) - \theta)]) \sin[2\theta]
\end{array}\right),              \label{coeff}
\eeq where $\phi(\theta)$ characterizes the light-hole wave and is
given by Eq.~(\ref{pphi}).

In the presence of the localized modes, by imposing the boundary
condition, $\psi_{\bf k}^{(H+)}(z=0) = 0$ we obtain for the
scattering coefficients in Eq.~(\ref{psihp2}) \beq
\left(\begin{array}{cccc} A_1(\theta) \\ A_2(\theta) \\ B_1(\theta)
\\ B_2(\theta)
\end{array}\right) =
\left(\begin{array}{cccc}
\frac{\left[-13 + \cos(2\theta) -\cos(2\chi) +  \cos(2\theta)\cos(2\chi) \right]\sin(\theta)}{5 + 9\cos(2\theta) + 5\cos(2\chi) + 3\cos(2\theta)\cos(2\chi) + 12i\sin(2\theta)\sin(\chi)} \\
\\
\frac{\left[-5 + \cos(2\theta) +7\cos(2\chi) +  \cos(2\theta)\cos(2\chi) \right]\cos(\theta)}{5 + 9\cos(2\theta) + 5\cos(2\chi) + 3\cos(2\theta)\cos(2\chi) + 12i\sin(2\theta)\sin(\chi)} \\
\\
\frac{-i\sqrt{3}\sqrt{3 - \cos(2 \chi) + 2 \sin(\chi)}\left[\cos(\frac{\theta}{2}) - i \sin(\frac{\theta}{2})\right]\sin(\theta)}{2\cos(\theta)(1+2\sin(\chi)) -2i\sin(\theta)(2+\sin(\chi))} \\
\\
\frac{-\sqrt{3}\sqrt{3 - \cos(2 \chi) - 2
\sin(\chi)}\left[\cos(\frac{\theta}{2}) + i
\sin(\frac{\theta}{2})\right]\sin(\theta)}{2\cos(\theta)(1-2\sin(\chi))
+2i\sin(\theta)(2-\sin(\chi))}
\end{array}\right).              \label{coeff2}
\eeq

\bibliography{edge}

\end{document}